\documentclass[10pt, conference, letter, twocolumn]{IEEEtran}
\pdfoutput=1
\usepackage{subfigure}
\usepackage{times}
\usepackage{latexsym}
\usepackage{morefloats}
\usepackage{bbold,bbm,dsfont}
\usepackage{amssymb}
\usepackage{amsmath}

\usepackage{enumitem}
\usepackage{enumerate}
\usepackage{mathrsfs}
\usepackage{amsgen,amsfonts,amsbsy,amsthm}
\usepackage{graphicx}
\usepackage{array}
\usepackage{hyperref}

\IEEEoverridecommandlockouts

\IEEEpubid{\makebox[\columnwidth]{} \hspace{\columnsep}\makebox[\columnwidth]{ }}

\begin{document}

\title{Performance Analysis of Beacon-Less \mbox{IEEE 802.15.4} Multi-Hop Networks}

\author{\IEEEauthorblockN{Rachit Srivastava}
		\IEEEauthorblockA{Deptt. of Electrical Communication Engineering \\
						Indian Institute of Science \\
						Bangalore, India 560012\\
						rachitsri@gmail.com}
\and
		\IEEEauthorblockN{Anurag Kumar}
		\IEEEauthorblockA{Deptt. of Electrical Communication Engineering \\
						Indian Institute of Science\\
						Bangalore, India 560012 \\
						anurag@ece.iisc.ernet.in}

\thanks{This work was supported by a research grant from the Department of Information Technology (DIT), Government of India, through the Automation Systems Technology (ASTEC) program. We are also thankful to Sanjay Motilal Ladwa for help in improving the results from Qualnet simulations.}
}

\maketitle

\begin{abstract}
	We develop an approximate analytical technique for evaluating the performance of multi-hop networks based on beacon-less CSMA/CA as standardised in IEEE 802.15.4, a popular standard for  wireless sensor networks. The network comprises sensor nodes, which generate measurement packets, and relay nodes which only forward packets. We consider a detailed stochastic process at each node, and analyse this process taking into account the interaction with neighbouring nodes via certain unknown variables (e.g., channel sensing rates, collision probabilities, etc.). By coupling these analyses of the various nodes, we obtain fixed point equations that can be solved numerically to obtain the unknown variables, thereby yielding approximations of time average performance measures, such as packet discard probabilities and average queueing delays. Different analyses arise for networks with no hidden nodes and networks with hidden nodes.  We apply this approach to the performance analysis of tree networks rooted at a data sink.  Finally, we provide a validation of our analysis technique against simulations.
\end{abstract}

\section{Introduction} \label{sec:introduction}

IEEE 802.15.4 \cite{winet.IEEE802-15-4-06std}, a popular standard for
the physical layer and medium access control for low-power wireless
networks, underlies the now well-known ZigBee standards for wireless
sensor networking. The standard defines two types of CSMA/CA
algorithms - slotted and unslotted, and is named on the basis of the
algorithm chosen - beaconed and beacon-less, respectively. With the
growing importance of wireless sensor networks in
industrial applications
\cite{winet.gungor-hancke09industrial-WSN-survey}, we need analysis
and design techniques for multi-hop IEEE 802.15.4 networks. In this
paper we develop a new approximate analytical technique for multi-hop
beacon-less IEEE~802.15.4 networks.

\begin{figure}
\begin{center}
  \includegraphics[scale=0.35]{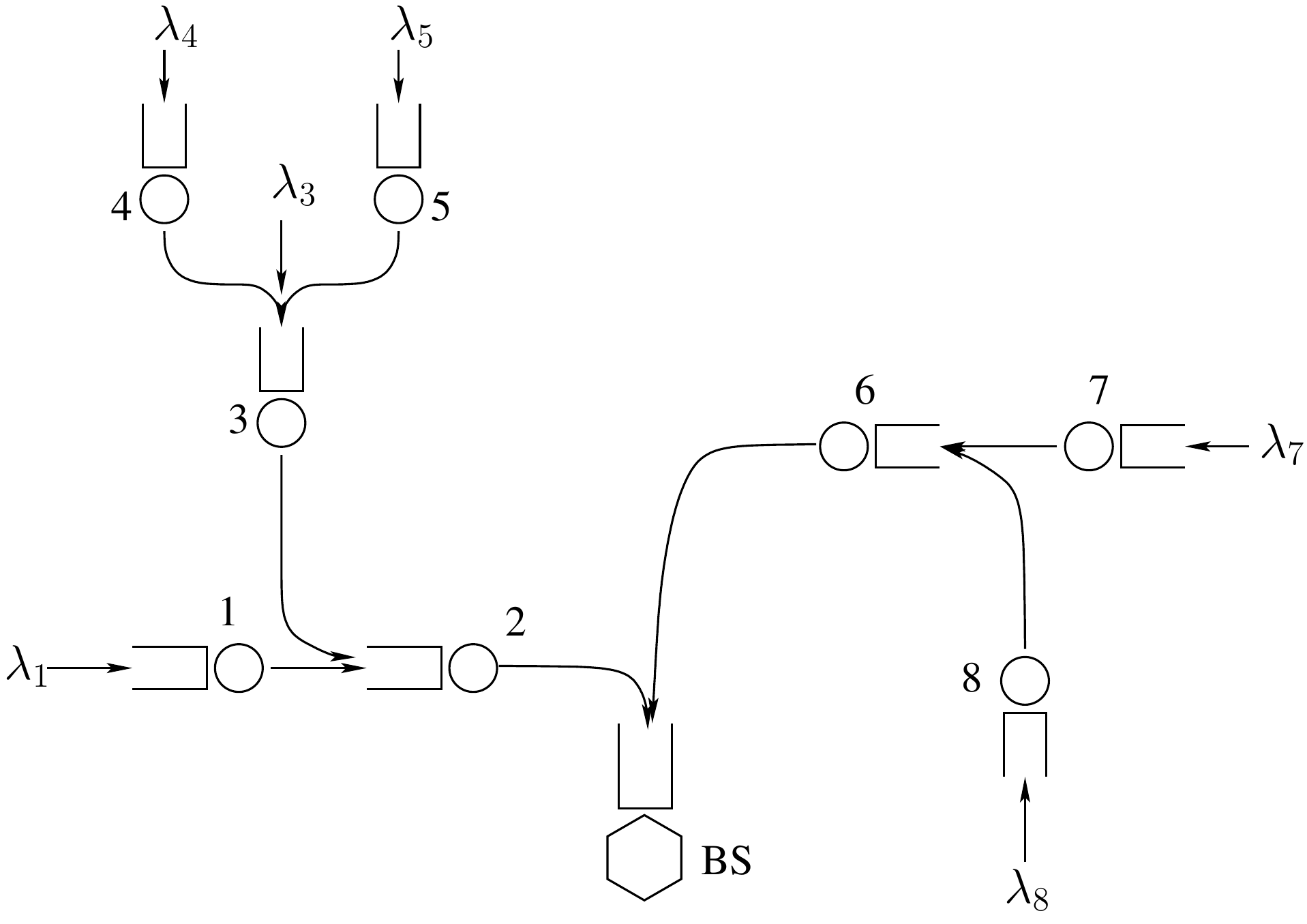}
  \caption{A WSN consisting of nodes arranged in a TREE
    topology. Source node $i$ has packet generation rate
    $\lambda_i$. Relay nodes $2$ and $6$ forward the data of their
    predecessors to the base-station BS.}
  \label{fig:network}
  \vspace{-7mm}
\end{center}
\end{figure}

In Figure~\ref{fig:network}, we depict the queueing schematic of a
network of the type that we are concerned with in this paper. There
are sensor nodes, namely, 1, 3, 4, 5, 7, and 8, that generate
measurements (at rate $\lambda_i$) that need to be transported to the
base station (BS).  The sensor nodes can also serve as \emph{relays}
for the traffic of other sensor nodes, e.g., Node~3 is a relay for
nodes 4 and 5.  Two additional nodes, 2 and 6, serve only as relay
nodes, due to the limited range of the radios associated with the
sensor nodes. The nodes use unslotted CSMA/CA, as standardised in
IEEE~802.15.4, to contend for the wireless medium and transmit their
packets. Even the stability analysis of such networks is a difficult
problem, due to the fact that the service rates applied to queues are
non-monotone with the queue occupancy of the other queues. In
principle, the entire network can be modeled via a large coupled
Markov chain, with the state at each node being the number of packets
in its queue, and the state of the contention process for the
head-of-the-line (HOL) packet. Such an approach is well known to be
intractable even for a network with a single contention domain and
saturated queues (see
\cite{winet.bianchi00performance,winet.kumar-etal04new-insights}). Thus
all researchers have taken recourse to developing approximate
analyses. Table~\ref{tbl:related-literature} summarizes some of the recent work on modeling approaches for CSMA/CA networks where we have listed the major limitations of the proposed models as well. Owing to the need for low power operation and large coverage, in general, a sensor network is multi-hop with the presence of hidden nodes. Further, for networks that carry measurement traffic, the queue occupancy varies with time. In addition, unsynchronized operation is preferred to save large overheads involved in the synchonizations of devices. Unlike other models, we consider all of these attributes in our modeling approach.

\scriptsize
\begin{table*}[ht]
  \centering
  \caption{A comparison of literature on analytical performance models of CSMA/CA networks.}
  \begin{tabular}{|p{23mm}|p{45mm}|p{100mm}|}
   \hline
   \textbf{Authors}  &  \textbf{Network Scenario \& Limitations}  & \textbf{Modeling Approach}  \\
   \hline
   Kim et al.\cite{winet.kim-etal06unslotted-CSMACA}
   	& 
	beacon-less 802.15.4; rare packet arrivals, star network; no hidden nodes; no queueing at nodes; with ACKs
   	& 
	2-dim discrete time Markov chain for each node with states: (backoff stage, backoff counter) while in backoff, and separate states when transmitting and idling; with unknown transition probabilities; coupling the per node models to get the unknowns via fixed point approach
	\\ \hline
   Singh et al.\cite{winet.singh-etal08slotted-zigbee-star}
   &
   beaconed 802.15.4; (un)saturated, star network; no hidden nodes; with ACKs; infinite queue size; homogeneous nodes
   &
   Markov renewal process at each node with cycle length dependent on the number of nodes available to attempt in the first backoff period of the cycle; transition probabilities and conditional expectations using node decoupling; coupling via stationary probabilites of remaining nodes; saturation analysis results used for finite arrivals
   \\ \hline
   Qiu et al.\cite{winet.qiu-etal07general-model-wireless-interference} 
   & 
   802.11 DCF; (un)saturated, star network; with hidden nodes and ACKs; infinite queue size
   & 
   discrete time Markov chain with states as the set of transmitting nodes having unknown transition probabilities, together with RSS measurements between each pair of node to obtain CCA and packet failure probabilities; iterative procedure for solving unknowns
   \\ \hline   
   He et al.\cite{winet.he-etal09slotted-CSMACA} &  beaconed 802.15.4; saturated, star network; no hidden nodes; with ACKs; homogeneous nodes
   & 
   two Markov chains for each node: one embedded at the end of transmissions in the common channel, and the other during backoffs embedded at each slot; both having states as (backoff stage, backoff counter); chains coupled to get a fixed point approach
   \\ \hline
   Martalo et al.\cite{winet.martalo-etal09slotted-CSMACA}  
   &
   beaconed 802.15.4; unsaturated, star network; no hidden nodes; no ACKs; finite queue size
   &
   two semi-Markov processes: one for tagged node having unknown transition probabilities, and the other for the shared radio channel that couples all the per node models to obtain the unknown quantities; MGFs used for finding queueing delays
   \\ \hline
   Goyal et al.\cite{winet.goyal-etal09beaconless-zigbee} 
   & 
   beacon-less 802.15.4; unsaturated star network; no hidden nodes; no queueing at nodes; with ACKs 
   &
   tracked the number of nodes with non-empty queues assuming that the probability of $m$ nodes having non-empty queues at any given time is same as the probability of $m-1$ empty nodes getting a new packet to send while a non-empty node is sending its current packet (for coupling individual node models); assumed that the CCA attempt processes of other nodes with non-empty queues are Poisson processes (decoupling)
   \\ \hline   
   Lauwens et al.\cite{winet.lauwens-etal09unslotted-CSMACA-analysis} 
   & 
   beacon-less 802.15.4; saturated, star network; no hidden nodes; no ACKs
   & 
   semi-Markov model for each node with unknown and non-homogeneous transition probabilities based on the backoff stage of the node; also found the distribution of backoff intervals and iterated over these together with the stationary probabilities of the semi-Markov processes
   \\ \hline
   DiMarco et al.\cite{winet.dimarco-etal10modeling-ieee-802-15-4-multihop} 
   & 
   beacon-less 802.15.4; unsaturated, multi-hop network; no queueing at nodes; with hidden nodes and ACKs 
   & 
   3-dim discrete time Markov chain for each node having state: (backoff stage, backoff counter, retransmission counter) with CCA and collision probabilities (unknown) as state transition probabilities (decoupling); assumed independence of node processes to find expressions for the CCA and collision probabilities using stationary probabilities of different Markov chains (coupling) 
   \\ \hline
   Shyam-Kumar\cite{winet.shyam-kumar10me-thesis} 
   & 
   beacon-less 802.15.4; (un)saturated, multi-hop network; with hidden nodes and ACKs; infinite queue size; mismatching results 
   & 
   saturated analysis using continuous time Markov chain with states as the set of transmitting nodes having unknown transition probabilities; found CCA and failure probabilities using steady state probability of the Markov chain; unsaturated network modeled as a Markov renewal process constituted by the nodes having non-empty queues, saturation analysis results used in each cycle with the set of non-empty nodes
   \\ \hline
  \end{tabular}
  \label{tbl:related-literature}
  \vspace{-5mm}
\end{table*}
\normalsize

\subsection*{Our Contribution} 
We consider a multi-hop WSN consisting of sources and relays arranged in a tree topology operating in the
beacon-less mode with ACKs. Each node has an infinite buffer space and
may operate in the saturated or unsaturated regime with fixed packet length. Different analysis techniques are
developed for networks with hidden nodes and networks devoid of hidden
nodes due to the difference in activity lengths perceived by a node in
the presence and absence of hidden nodes. We model the stochastic process evolving at a node by
incorporating the influence of the other node in the network by their
(unknown) time averaged statistics. After making certain
approximations, we use renewal theoretic arguments to couple
individual node processes and obtain the unknown variables as fixed
points of an iterative scheme. Although this \emph{decoupling} (or \emph{mean-field}) approximation is popular in such situations, our more detailed model incorporating several issues not considered together hitherto requires a careful handling of the analysis. We identify and calculate two QoS
measures for each source node, viz., the packet delivery probability and
end-to-end packet delay, in terms of these variables. We observe that
in a multi-hop network, the packet delivery probability falls sharply
before the end-to-end packet delay becomes substantial.

The rest of the paper is organized as follows: In Section~\ref{sec:overview-of-standard}, we give an overview of the
unslotted CSMA/CA mechanism. Section~\ref{sec:analytical-model}
explains the node behaviour and elaborate upon the different
analysis techniques used for networks without hidden nodes and
networks with hidden
nodes. Section~\ref{sec:numerical-and-simulation-results} compares our
analysis with simulations. We conclude the paper in
Section~\ref{sec:conclusion}.

\section{Unslotted CSMA/CA for Beacon-less IEEE 802.15.4 Networks}\label{sec:overview-of-standard}

Node behaviour under unslotted CSMA/CA is shown in Figure~\ref{fig:evolution}. A node with an empty queue remains idle until it generates a packet or receives one from its predecessor nodes. When a node has some data to send (non-empty queue), it initiates a random backoff. The first backoff period is sampled uniformly from $0$ to \emph{${2^{macminBE}-1}$}, where \emph{macminBE} is a parameter fixed by the standard. The backoff period is specified in terms of slots where a slot equals 20 symbol times ($T_s$) and a symbol time equals 16~$\mu$s. The node then performs a CCA (\emph{clear channel assessment}) to determine whether the channel is idle. If the CCA succeeds, the node does a \emph{Rx-to-Tx turnaround}, which is worth $12$ symbol times and starts transmitting on the channel. The failure of the CCA starts a new backoff process with the backoff exponent raised by one, i.e., to \emph{macminBE+1}, provided it is lesser than the maximum backoff value given by \emph{macmaxBE}. The maximum number of successive CCA failures for the same packet is governed by \emph{macMaxCSMABackoffs}, exceeding which the packet is discarded at the MAC layer. The standard allows the inclusion of acknowledgements (ACKs) which are sent by the intended receivers on a successful packet reception. Once the packet is received, the receiver performs a \emph{Rx-to-Tx turnaround}, which is again $12$ symbol times, and sends a $22$ symbol fixed size ACK packet. A successful transmission is followed by an \emph{InterFrame Spacing}(IFS) before sending another packet.

When a transmitted packet collides or is corrupted by the PHY layer noise, the ACK packet is not generated which is interpreted by the transmitter as a failure delivery. The node retransmits the same packet for a maximum of \emph{aMaxFrameRetries} times before discarding it at the MAC layer. After transmitting a packet, the node turns to Rx-mode and waits for the ACK.  The \emph{macAckWaitDuration} determines the maximum amount of time a node must wait for in order to receive the ACK before concluding that the packet (or the ACK) has collided. The default values of \emph{macminBE}, \emph{macmaxBE}, \emph{macMaxCSMABackoffs} and \emph{aMaxFrameRetries} are 3,5,4 and 3, respectively.

\section{Analytical Model} \label{sec:analytical-model}

\begin{figure}[ht]
	\begin{center}
		\includegraphics[height=4cm, width=9cm]{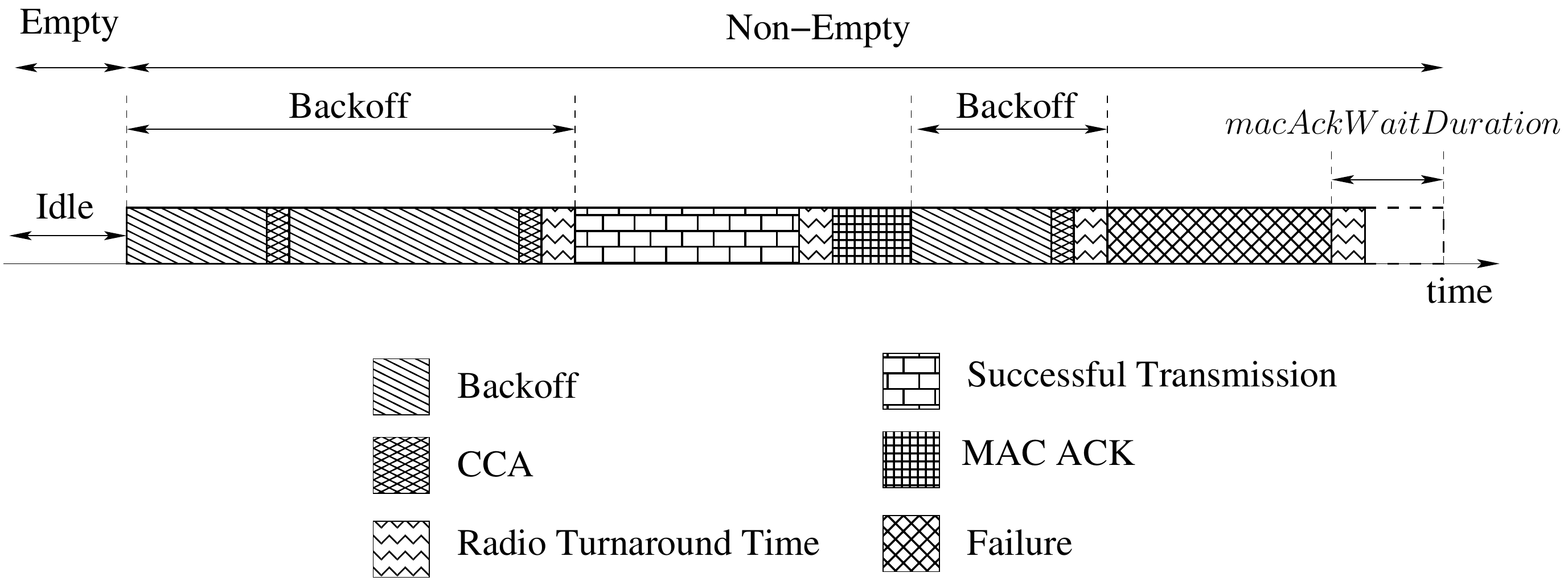}
		\caption{Process evolving at Node~$i$ where it goes through various states.}
		\label{fig:evolution}
		\vspace{-5mm}
	\end{center}
\end{figure}

As shown in Figure~\ref{fig:network}, the network consists of sensors, relays and a base-station (BS). We consider time invariant links and static (immobile) nodes. We assume that the sensors generate traffic according to independent point processes. These constitute the aggregate external arrival process for the sensor network. Each sensor transmits its data to the next-hop node according to the topology (Tree or Star). The intermediate nodes along a route may be relays in which case they simply forward the incoming traffic, or they may be sensors which transmit their own packets as well as the received packets. Based on the network congestion, the nodes may discard packets due to consecutive failed CCAs or frame retries. Figure~\ref{fig:evolution} shows a typical sample path of the process evolution at a node $i$. 

\begin{figure}[htbp]
	\begin{center}
		\includegraphics[scale=0.45]{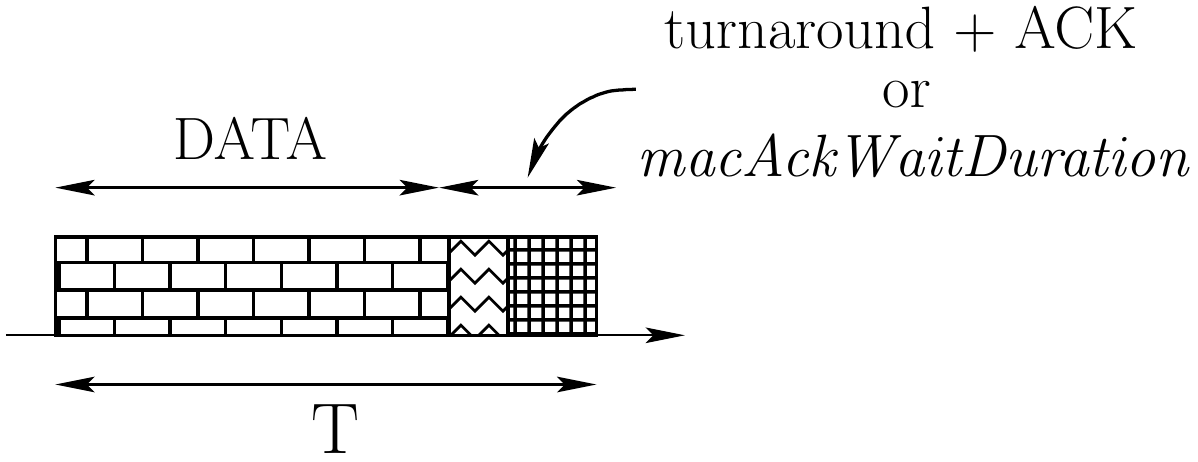}
		\caption{The transmission period $T$ which includes the DATA transmission time $T_x$, the turnaround time of $12~T_s$, and the MAC ACK duration of $22~T_s$ (or the \emph{macAckWaitDuration} of $34~T_s$, by \textbf{(S8)}).}	
		\label{fig:transmission}
		\vspace{-5mm}
	\end{center}
\end{figure}

Since the process evolutions at various nodes are coupled and each node (source or relay) can be present in any state, the random process representing the system is multi-dimensional and non-homogeneous. Further, different nodes can interact with different subsets of nodes based on their Carrier Sense (CS) range and positions, the process is also non-symmetric. The exact analysis of such a process is intractable. We, therefore, make several simplifications:
\begin{description}[leftmargin=0cm]
\item [(S1)]All packets are of the same fixed length. \emph{Hence, the DATA transmission duration is fixed.}
\item [(S2)]The internode propagation delays are of the order of nanoseconds, and hence can be taken to be zero.
\item [(S3)]Zero packet capture. \emph{As a result, all the packets involved in a collision get corrupted.}
\item [(S4)]The IFS is neglected. \emph{It means that the receiver does not discard the packet received during the IFS.}
\item [(S5)]The packet error probability at each link is fixed and known. \emph{As already mentioned, the links are not time varying and we can empirically obtain the error characteristics of a link.}
\item [(S6)]ACK packets are short and, therefore, not corrupted by PHY layer noise. 
\item[(S7)]A node's CCA succeeds when there is no transmission by any node in its CS range at the time of initiation of its CCA. \emph{Recall that in the standard, the channel state is averaged over the $8$~symbol duration.}
\item [(S8)]The time taken by a transmitting node (called the transmission period and denoted by $T$) for the activities of successful transmission and collision are the same. \emph{If the transmitted data collides at the receiver, the macAckWaitDuration for the transmitting node is equal to the sum of the turnaround time of $12~T_s$ and the ACK duration of $22~T_s$, a total of $34~T_s$ (see Figure~\ref{fig:transmission}).}
\item [(S9)]All nodes have the same CS and transmission ranges. \emph{This implies that if a node $i$ is in CS (resp. transmission) range of another node $j$, then so is node $j$ in the CS (resp. transmission) range of node $i$.}
\end{description}

Further, we make use of a \emph{decoupling approximation} whereby we model each node separately, incorporating the influence of the other nodes in the network by their average statistics and as if these nodes were independent of the tagged node. 

\begin{figure}[h]
\begin{center}
\includegraphics[width=9cm]{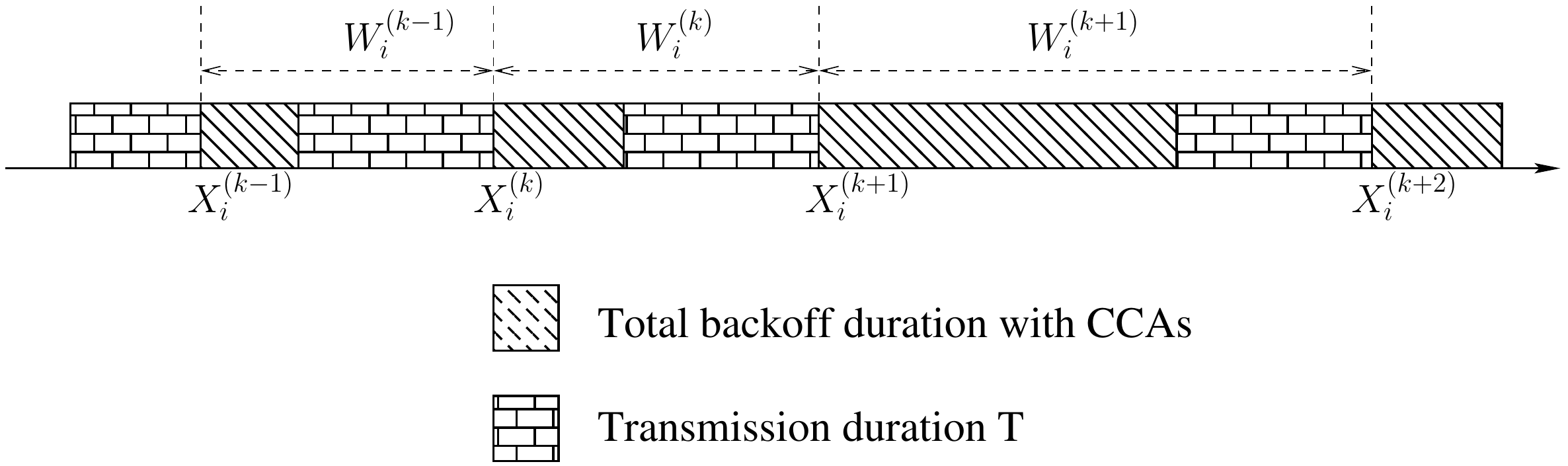}
\caption{The renewal process obtained by observing the process at a node $i$ after removing all the idle time periods from the original process shown in Figure~\ref{fig:evolution}. The renewal epochs are denoted by $\{X_i^{(k)}\}$ and the cycle lengths by $\{W_i^{(k)}\}$.}
\label{fig:ARP_calc_of_rates}
\vspace{-5mm}
\end{center}
\end{figure}

The channel activity perceived by any node $i$ is only due to nodes that are within its CS range (denoted by $\Omega_i$). Let us consider the durations during which the queue at node $i$ is non-empty. During these times, node $i$ alternates between performing CCAs, and transmitting the packets from its queue. Let the CCA attempt process at node $i$ be a Poisson process of rate $\beta_i$ conditioned on being in backoff periods. Let $\overline{\beta}_j$ be the CCA attempt rate of node $j$ in $\Omega_i$ over all time, and approximate the CCA attempt processes for the nodes in $\Omega_i$ by independent Poisson processes. By modeling simplification \textbf{(S1)}, we assume that all packets entering node $i$ have the same fixed length, and hence take the same amount of time when being transmitted over the medium (denoted by $T$). As a result, we observe that as perceived by node $i$ in conditional time, its transmission periods and the intervening backoff periods constitute a renewal process, as shown in Figure~\ref{fig:ARP_calc_of_rates}.
\begin{figure}[h]
	\begin{center}
		\includegraphics[scale=0.4]{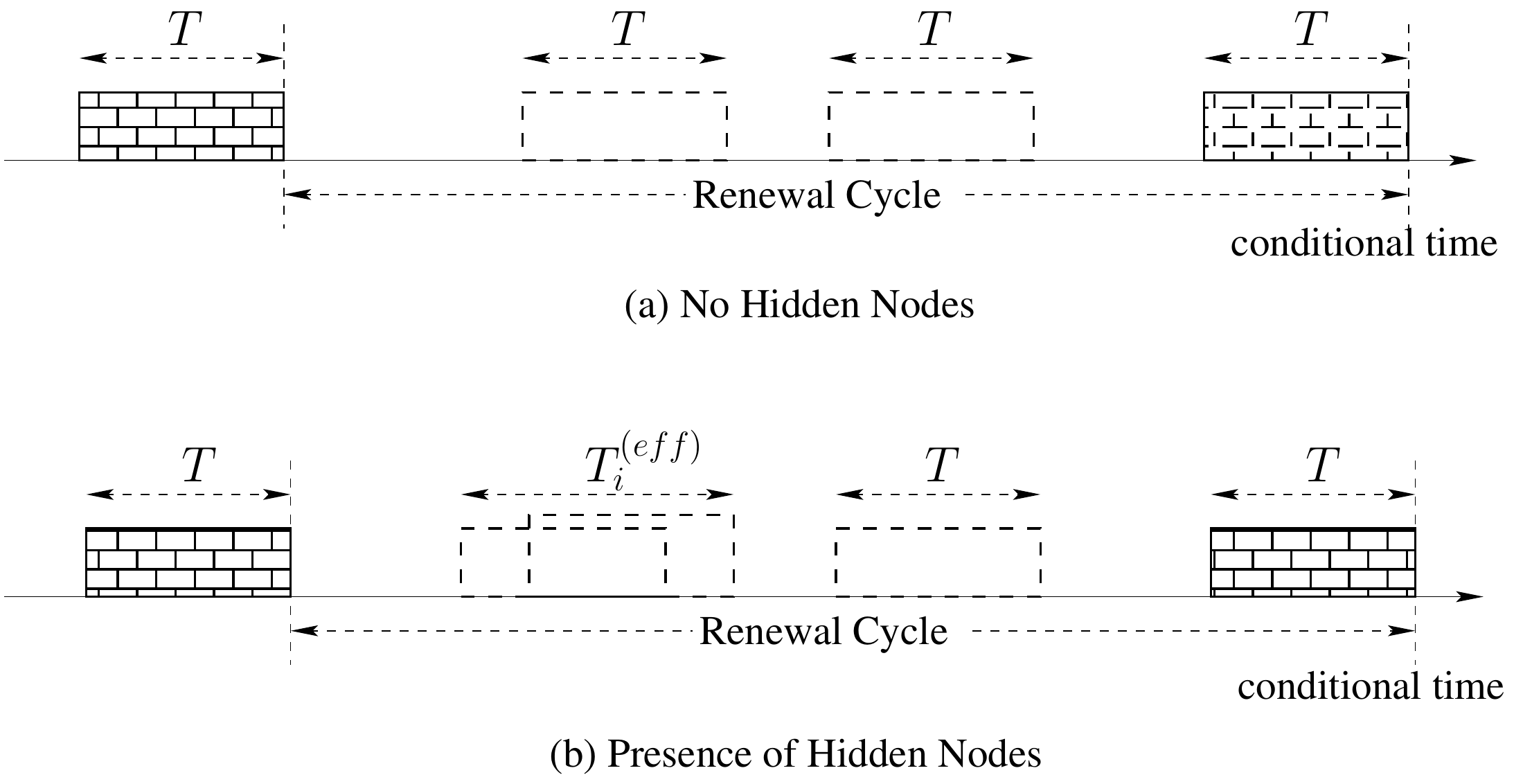}
		\caption{(a) expands a non-empty period (\emph{conditional time}) and displays a renewal cycle when there are no hidden nodes in the system, and (b) shows the expanded non-empty period when hidden nodes are present.}
		\label{fig:system}
		\vspace{-5mm}
	\end{center}
\end{figure}
Note that $\overline{\beta}_j$ is equal to the product $\beta_jb_jq_j$ where $\beta_j$ is the CCA attempt rate of node $j$ in backoff periods, $b_j$ is the fraction of time node $j$ is in backoff provided its queue is non-empty, and $q_j$ is the fraction of time node $j$'s queue is non-empty. Figure~\ref{fig:system} shows a renewal cycle of node $i$. The dashed transmission durations belong to the nodes in $\Omega_i$.

\begin{figure}[h]
\begin{center}
	\includegraphics[scale=0.45]{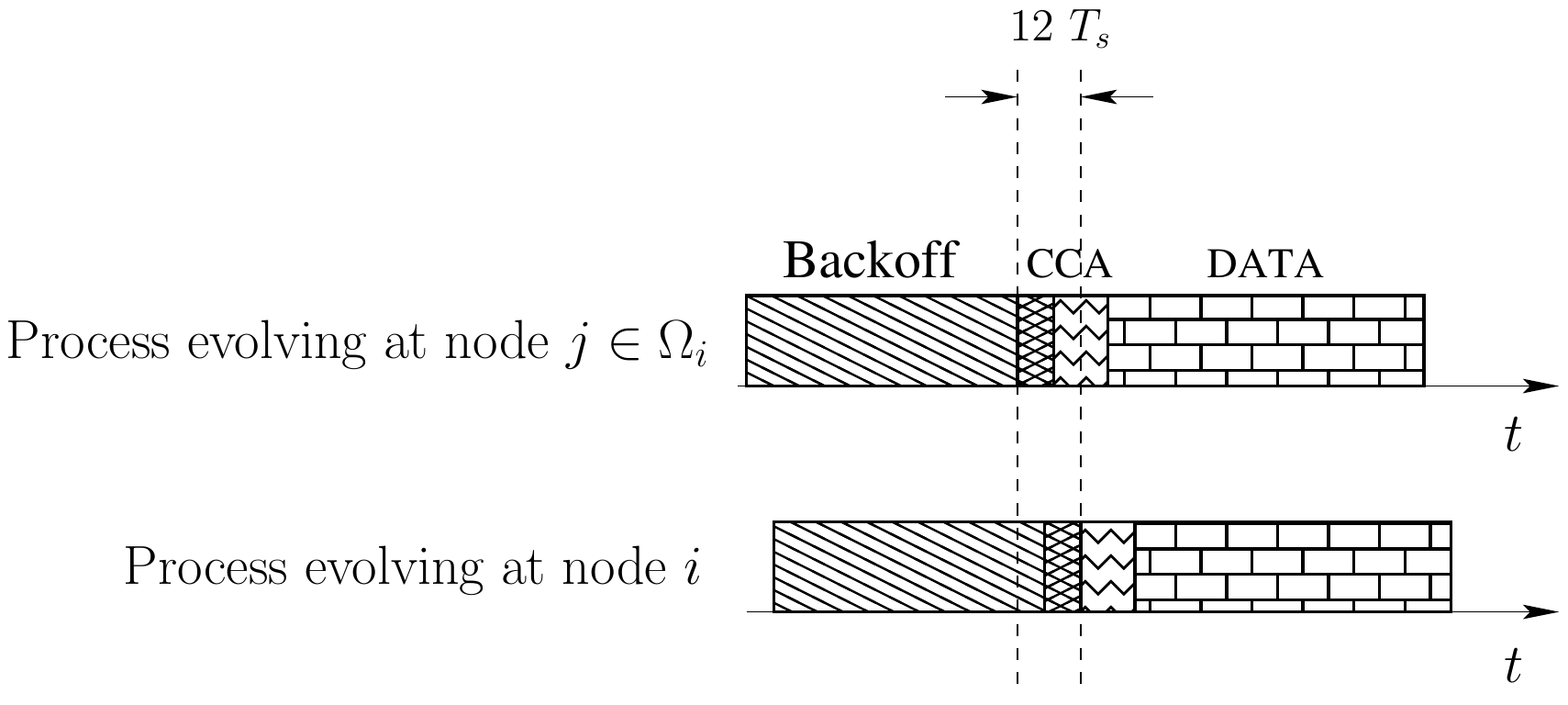}
	\caption{Node $j\in \Omega_i$ finishes its backoff, performs a CCA, finds the channel idle and starts transmitting the DATA packet. Node $i$ finishes its backoff anywhere in the shown $12~T_s$ duration and there is no other ongoing transmission in $\Omega_i$, its CCA succeeds and it enters the transmission duration. As a result, the DATA packets may collide at $r(i)\in \Omega_j$ and/or $r(j) \in \Omega_i$.}
	\label{fig:simul_channel sensing}
	\vspace{-5mm}
\end{center}
\end{figure}

\begin{figure}[h]
\begin{center}
\includegraphics[scale=0.45]{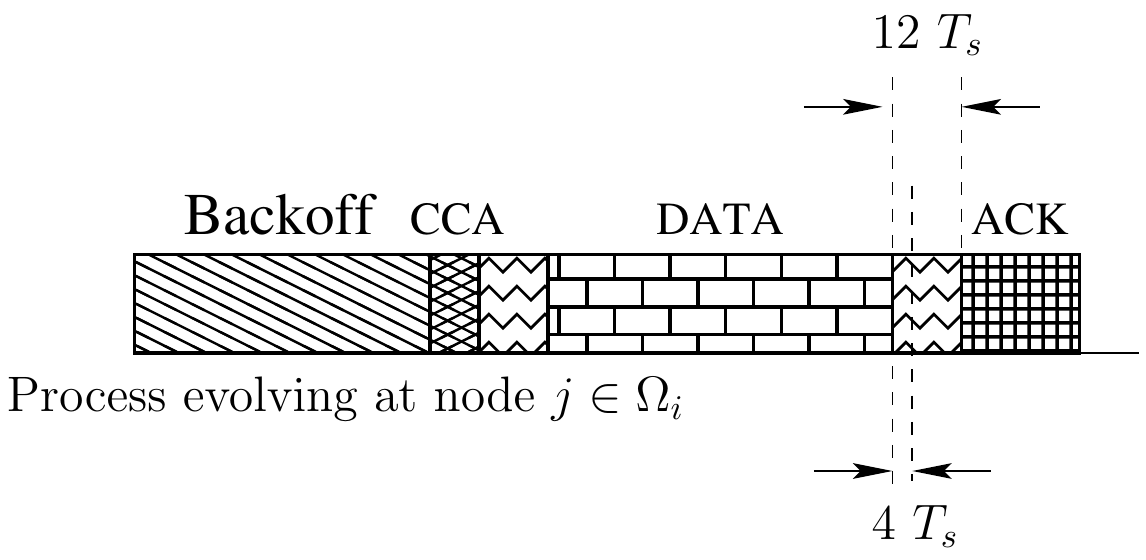}
	\caption{Vulnerable window of $4~T_s$ in the transmission period of node $j\in \Omega_i$ during which node $i$'s CCA attempt would be successful.}
	\label{fig:CCAfail}
	\vspace{-7mm}
\end{center}
\end{figure}

Before discussing the difference in activity lengths in the two network types (as shown in Figure~\ref{fig:system}), we list the various packet collision scenarios. Firsty, the majority of packet collisions in a wireless network occur due to the presence of hidden nodes. However, there are scenarios where transmissions from two transmitters placed within the CS range of each other can overlap, known as \emph{Simultaneous Channel Sensing} and \emph{Vulnerable Window} and are explained in Figures~\ref{fig:simul_channel sensing} and \ref{fig:CCAfail}, respectively. Note that we are making use of \textbf{(S2)} to avoid including insignificant timing details here.

Further, the ACK packet size is just a fraction of DATA packet size (e.g. compare an ACK packet ($22$~symbols) with a DATA packet of length $260$~symbols at PHY layer). It means that if a node is in its transmission period $T$, then we can say with high probability that it would be in the $T_x$ (DATA) duration. This fact further says that the probability of a node's CCA initiating in the vulnerable window of another node is negligible. Hence, we can safely assume that the CCAs seldom fail due to ACKs. Similarly, the small ACK packets can rarely cause packet collisions.

We now return to elucidate the difference in length of activity periods in a renewal cycle.
\begin{description}[leftmargin=0cm]
\item[(a) Absence of Hidden Nodes:] When a node transmits, all the other nodes in the network can hear it resulting in CCA failures for other nodes that try to assess the channel in the transmission period. Also if two nodes are involved in simultaneous channel sensing, the activity period may extend from $T$ to a maximum of $(T+12~T_s)$. Since $12~T_s<<T$, we can assume that the activity period is only a single transmission period of duration $T$. 
\item[(b) Presence of Hidden Nodes:] Since a node's hearing capacity is limited, it may not perceive the activities of all the nodes in the network which may cause dilation of activity period.
\begin{figure}[h]
\begin{center}
		\includegraphics[scale=0.3]{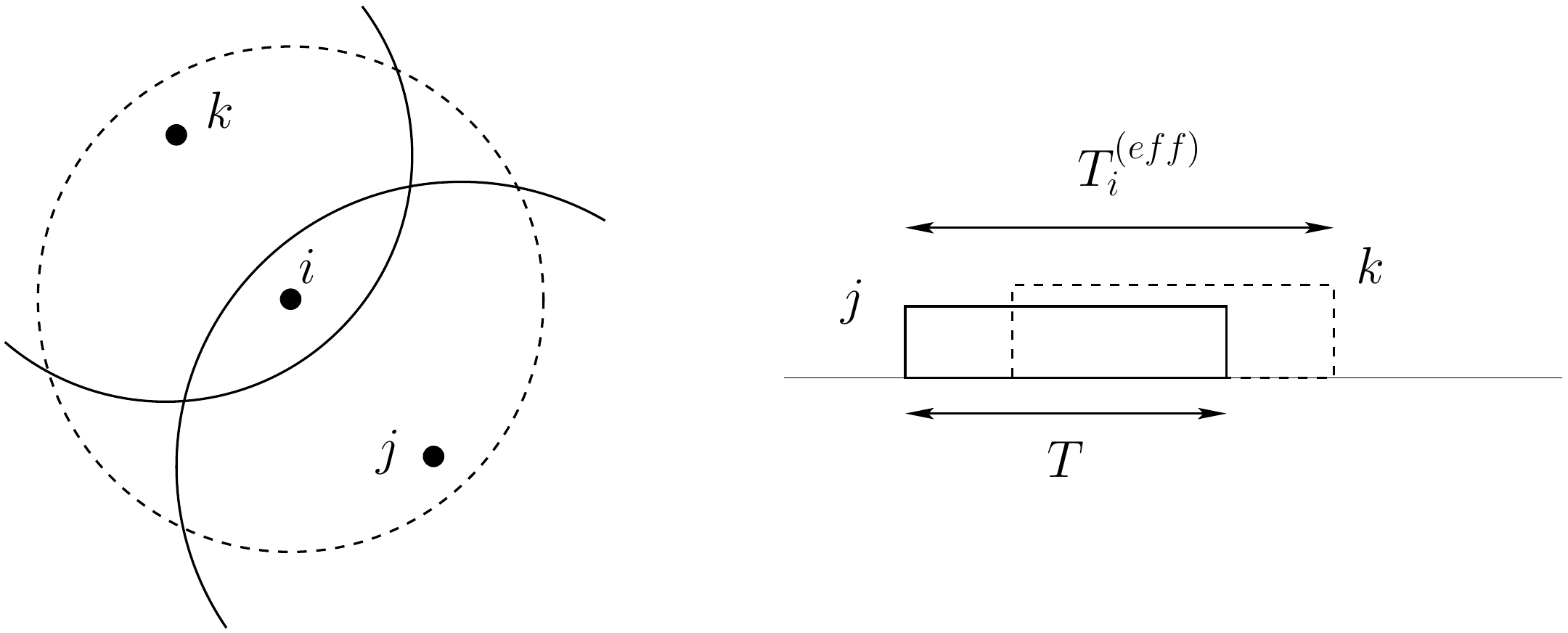}
		\caption{\emph{Dilation of transmission period as perceived by node $i$}.}
		\label{fig:t_eff}
		\vspace{-5mm}
\end{center}
\end{figure}
As shown in Figure~\ref{fig:t_eff}, nodes $j$ and $k$ are hidden from each other with respect to the receiver node $i$. Let node $j$ starts transmitting while nodes $i$ and $k$ are silent. Since node $k$ cannot hear node $j$, it can start its own transmission while the transmission from node $j$ is still going on. For node $i$, this looks like a dilated transmission period, whose length is denoted by $T_i^{(eff)}$. 
\end{description}

Since each node has a different set of neighbours, we make the following approximation to simplify our analysis: \emph{The mean length of Dilated Activity Period as perceived by node $i$ is equal to the length of mean busy period of an $M/D/\infty$ queue where the deterministic service time is equal to a single transmission period $T$, and the arrival process is approximated as a Poisson process having rate equal to the aggregate transmission initiation rate of all the nodes in the CS range of node $i$}. Note that the above assumption is equivalent to saying that all nodes in $\Omega_i$ are hidden from each other, and hence results in a larger mean for the dilated activity period.

Let $\overline{\tau}_j$ be the rate of successful CCAs by node $j$, or equivalently, the rate of transmission initiations by node $j$, which is equal to the product $\overline{\beta}_j(1-\alpha_j)$ where $\alpha_j$ is the probability of CCA failure for node $j$. For any node $i$, let the aggregate transmission initiation rate for nodes in $\Omega_i$ is
\begin{eqnarray}
\zeta_i = \sum_{j\in \Omega_i}\overline{\tau}_j 
\label{eqn:zeta}
\end{eqnarray}
Assuming node $j\in \Omega_i$ has started transmission and another node $k\in \Omega_i$ starts at time $u<T$, the expression for $T_i^{(eff)}$ can be written recursively as
\begin{eqnarray}
\textstyle{T_i^{(eff)}} &\textstyle{=}& \textstyle{Te^{-\zeta_i T} + \int_0^T(u+T_i^{(eff)})\zeta_ie^{-\zeta_i u }du} \nonumber \\
&\textstyle{=}& \textstyle{\frac{1}{\zeta_i}\big(e^{\zeta_i T} - 1\big)}
\end{eqnarray}
\subsection{Topologies with No Hidden Nodes}
By the assumptions made so far, we have a superposition of independent Poisson processes during the backoff periods of node $i$. Define the following:
\begin{eqnarray*}
\eta_i = \frac{\beta_i}{\beta_i + \displaystyle{\sum_{j\neq i}\overline{\beta}_j}} \quad &;& \quad  g_i = \frac{1}{\beta_i + \displaystyle{\sum_{j\neq i}\overline{\beta}_j}} \\
\text{and} \qquad c_i &=& \Bigg(1-e^{-12~T_s\beta_i}\Bigg) 
\end{eqnarray*}
Since ZigBee is designed to operate at fairly low packet generation rates at which packet discard probability (due to successive CCAs or frame retries) is very low, all the recursive equations written below do not take packet discards into account.
\subsubsection{Probability of CCA Failure, $\alpha_i$}
The probability of CCA failure is the probability of the occurrence of at least one transmitting node in the CS range of node $i$, given that node $i$ performs a CCA. Defining $N_i(t)$ and $N_i^{(F)}(t)$ as the total number of CCAs and number of failed CCAs in the interval $(0,t]$, respectively, we have:
\begin{eqnarray*}
\alpha_i = \lim_{t \to \infty}\frac{N_i^{(F)}(t)}{N_i(t)} = \lim_{t \to \infty}\frac{\frac{N_i^{(F)}(t)}{t}}{\frac{N_i(t)}{t}}
\end{eqnarray*}
Applying Renewal-Reward Theorem (RRT) \cite{kulkarni95modeling-stochastic-systems} here, we get
\begin{eqnarray}
\alpha_i &=&\frac{\frac{N_i^{(F)}}{W_i}}{\frac{N_i}{W_i}}
\label{eqn:alpha_rates}
\end{eqnarray}
where $N_i^{(F)}$ is the mean number of failed CCAs and $N_i$ is the mean number of total CCAs in a cycle. The mean renewal cycle length, $W_i$, is given by
\begin{eqnarray*}
W_i &=& g_i+\eta_i T+(1-\eta_i)c_i T+(1-\eta_i)(1-c_i)(T+W_i)
\end{eqnarray*}
By taking the reward to be the mean number of total CCAs by node $i$ in a renewal cycle, i.e., $N_i$, and applying RRT, we can write
\begin{eqnarray*}
N_i &=& \eta_i + (1-\eta_i)c_i + (1-\eta_i)(1-c_i)\Big(\beta_i T + N_i\Big)
\end{eqnarray*}
Note that node $i$ performs CCAs at the rate of $\beta_i$ for the entire transmission period $T$ of another node, i.e., a total of $\beta_i T$ CCAs all of which fail and the cycle continues after this period. Again, let the reward be the total number of failed CCA attempts by node $i$ in a renewal cycle, say, $N_i^{(F)}$, which can be written using \textbf{(S7)} as
\begin{eqnarray*}
N_i^{(F)} &=& (1-\eta_i)(1-c_i)\Big(\beta_i T+N_i^{(F)}\Big)
\end{eqnarray*}
On rearranging, the CCA failure probability is given by
\begin{eqnarray}
\alpha_i &=& \frac{(1-\eta_i)(1-c_i)\beta_i T}{\eta_i + (1-\eta_i)c_i + (1-\eta_i)(1-c_i)\beta_i T}
\label{eqn:alpha}
\end{eqnarray}

\subsubsection{Packet Failure Probability, $\gamma_i$}
\begin{figure}[ht]
\begin{center}
	\includegraphics[scale=0.45]{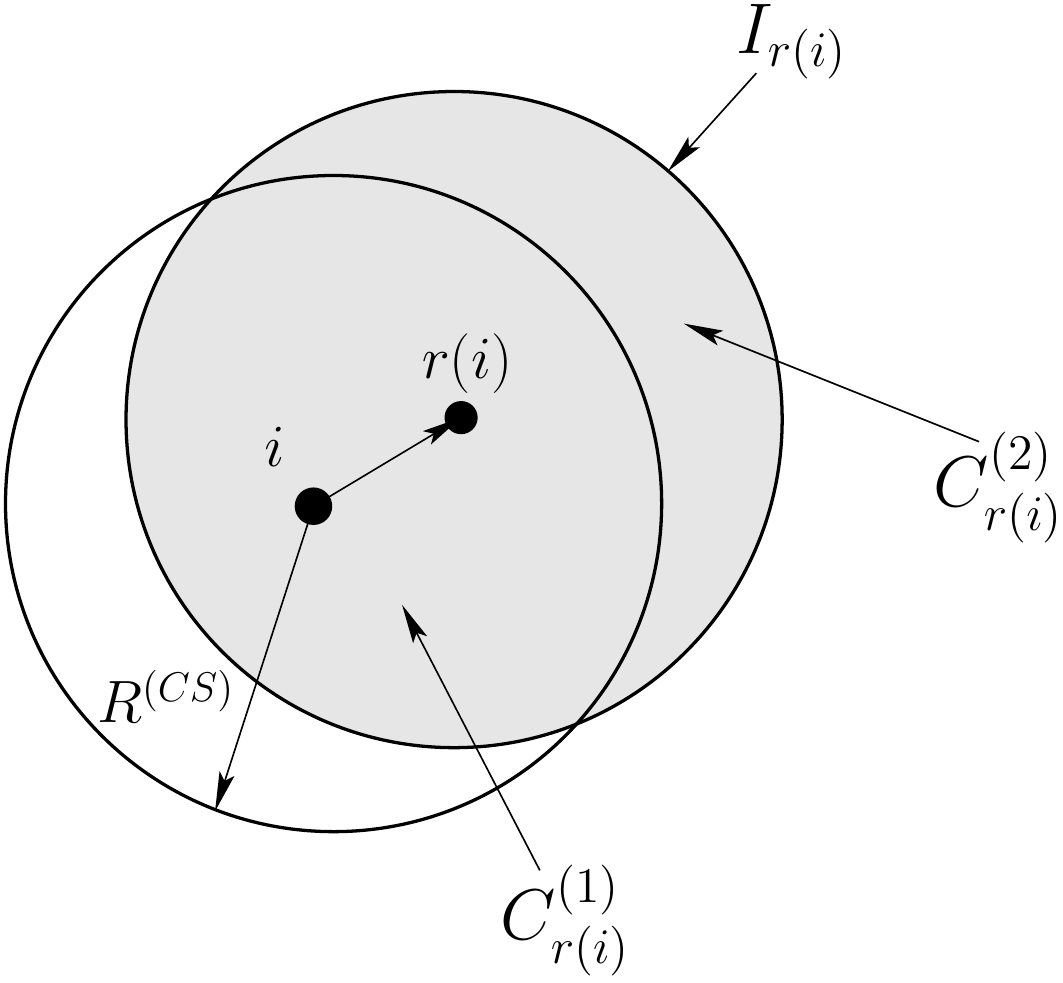}
    \caption{Interference region around the receiver. $R^{(CS)}$ is the CS range which is assumed to be equal for all nodes by \textbf{(S9)}.}
    \label{fig:interference}
	\vspace{-5mm}
\end{center}
\end{figure}

A transmitted packet can fail to be decoded by its intended receiver due to a collision, or due to noise. We do not take packet capture into account here (see \textbf{(S3)}). The probability of packet failure, $\gamma_i$, is defined as:
\begin{eqnarray}
\gamma_i &=& p_i +(1-p_i )l_i
\label{eqn:gamma}
\end{eqnarray}
where $p_i $ is the probability of packet collision and $l_i$ is the probability of \emph{data} packet error (known by \textbf{(S5)}) on the link between $i$ and its receiver $r(i)$ due to noise. Recall that ACKs are not corrupted by PHY noise (see \textbf{(S6)}). Defining $M_i(t)$ and $M_i^{(C)}(t)$ as the total number of transmissions and total number of collisions in interval $(0,t]$, we can write 
\begin{eqnarray*}
\gamma_i&=&\Bigg(\lim_{t \to \infty}\frac{M_i^{(C)}(t)}{M_i(t)}\Bigg)+\Bigg(1-\lim_{t \to \infty}\frac{M_i^{(C)}(t)}{M_i(t)}\Bigg)l_i
\end{eqnarray*}
Dividing each term by \emph{t} and applying RRT here as before, we get
\begin{eqnarray*}
\gamma_i &=& \Bigg(\frac{\frac{M_i^{(C)}}{W_i}}{\frac{M_i}{W_i}}\Bigg) + \Bigg(1 - \frac{\frac{M_i^{(C)}}{W_i}}{\frac{M_i}{W_i}}\Bigg)l_i
\end{eqnarray*}
where $M_i^{(C)}$ is the mean number of collided packets and $M_i$ is the mean number of transmitted packets in a cycle. If we ignore packet discards, there is a single transmission by node $i$ in each renewal cycle, $M_i$ becomes unity. We denote the set of nodes that can cause interference in successful reception at $r(i)$ by $I_{r(i)}$, which is composed of two sets:
\begin{center}
$C^{(1)}_{r(i)}=\{j\in\mathscr{N}: j\in\Omega_i\text{ and }j\in I_{r(i)}\}$\\
$C^{(2)}_{r(i)}=\{j\in\mathscr{N}: j \notin\Omega_i\text{ and }j \in I_{r(i)}\}$
\end{center}
such that
\begin{eqnarray*}
C^{(1)}_{r(i)}\cap C^{(2)}_{r(i)}&=&\emptyset\\ 
C^{(1)}_{r(i)}\cup C^{(2)}_{r(i)}&=&I_{r(i)} 
\end{eqnarray*}
Note that the receiver $r(i)$ is in $C^{(1)}_{r(i)}$ (see Figure~\ref{fig:interference}).
\vspace{-4mm}

In this case, the collisions can happen only due to the simultaneous channel sensing event. Let the reward be the mean number of collisions of node $i$'s packets in a renewal cycle, say, $M_i^{(C)}$, which can be written as
\begin{eqnarray*}
M_i^{(C)} &=& \eta_i\Big(1-\exp\{-12~T_s\displaystyle{\sum_{j\ne i}\overline{\beta}_j}\}\Big) + (1-\eta_i)c_i \\
 && + (1-\eta_i)(1-c_i)M_i^{(C)}
\end{eqnarray*}
Note that the first two terms are simultaneous channel sensing events. On rearranging ,the packet collision probability at node $i$ is given by
\begin{eqnarray}
p_i &=& \frac{\eta_i\Big(1-\exp\{-12~T_s\displaystyle{\sum_{j\ne i}\overline{\beta}_j}\}\Big) + (1-\eta_i)c_i}{1-(1-\eta_i)(1-c_i)}
\label{eqn:p}
\end{eqnarray}
The packet failure probability can then be calculated using Equations~\ref{eqn:gamma} and \ref{eqn:p}.

\subsection{Topologies with Hidden Nodes}
For each node $i$, we now need to keep track of channel activity in its CS range only. Further, even if a node $j \in \Omega_i$ finishes its backoff first when there is no activity in $\Omega_i$, its CCA need not be successful because it may be blocked by another node in the system hidden to $i$. Hence, we redefine $\eta_i$ and $g_i$ as
\begin{eqnarray*}
\eta_i &=& \frac{\beta_i}{\beta_i + \displaystyle{\sum_{j\in \Omega_i}\overline{\tau}_j}} \qquad \text{and} \qquad 
g_i = \frac{1}{\beta_i + \displaystyle{\sum_{j\in \Omega_i}\overline{\tau}_j}}
\end{eqnarray*} 
The expression for $c_i$ remains the same. 

\subsubsection{Probability of CCA Failure, $\alpha_i$}
The mean renewal cycle length $W_i$ can be written as
\begin{eqnarray*}
W_i=g_i+\eta_i T+(1-\eta_i)c_i T+(1-\eta_i)(1-c_i)(T_i^{(eff)}+W_i)
\end{eqnarray*}
Note that we have replaced $T$ in the last term by $T_i^{(eff)}$ because this period is susceptible to dilation. 

The probability of CCA failure, $\alpha_i$, can be evaluated on similar lines as before with the exception that $T$ gets replaced by $T_i^{(eff)}$, i.e.,
\begin{eqnarray}
\alpha_i &=& \frac{(1-\eta_i)(1-c_i)\beta_i T_i^{(eff)}}{\eta_i+(1-\eta_i)\Big((1-c_i)\beta_i T_i^{(eff)} + c_i\Big)}
\label{eqn:alpha_hidden}
\end{eqnarray}

\subsubsection{Packet Failure Probability, $\gamma_i$}

We first need to find the probability that nodes in a given set are not transmitting using a product approximation due to the unavailability of joint distribution of processes. Let $h_i$ be the fraction of time a node is not transmitting while its queue is non-empty. The reward $H_i$ in this case is the mean amount of time node $i$ is not transmitting in a renewal cycle given by
\begin{eqnarray*}
H_i &=& \frac{g_i+(1-\eta_i)(1-c_i)T_i^{(eff)}}{1-(1-\eta_i)(1-c_i)} 
\end{eqnarray*}
which in turn provides
\begin{eqnarray*}
h_i = \frac{H_i}{W_i} = \frac{g_i + (1-\eta_i)(1-c_i)T_i^{(eff)}}{g_i + \eta_i T + (1-\eta_i)c_iT + (1-\eta_i)(1-c_i)T_i^{(eff)}} 
\end{eqnarray*}
The fraction of time node $i$ is not transmitting (unconditional) is denoted by $\overline{h}_i$ and is equal to 
\begin{eqnarray}
\overline{h}_i &=& (1-q_i) + q_ih_i 
\end{eqnarray}

The expression for collision probability can be found as follows:
\begin{description}[leftmargin=0cm]
\item[1.] The first term accounts for the fact that Node $i$ started transmitting in the presence of at least one transmission by the hidden nodes set, i.e., $C^{(2)}_{r(i)}$. 
\begin{eqnarray*}
R_i^{(1)} &=& \eta_i\Bigg(1-\prod_{j\in C^{(2)}_{r(i)}}\overline{h}_j\Bigg)
\end{eqnarray*}
\item[2.] The second term accounts for the scenario when Node $i$ started its transmission as a simultaneous channel sensing with a node in $\Omega_i$ in the presence of at least one transmission by the hidden nodes set, i.e., $C^{(2)}_{r(i)}$.. 
\begin{eqnarray*}
R_i^{(2)} &=& (1-\eta_i)c_i\Bigg(1-\prod_{j\in C^{(2)}_{r(i)}}\overline{h}_j\Bigg)
\end{eqnarray*}
\item[3.] The third term considers the case where Node $i$ starts transmitting in the absence of any ongoing transmission by a hidden node but it encounters a simultaneous transmission by a node in $C^{(1)}_{r(i)}$ anywhere in the corresponding $12~T_s$ period and/or a transmission by a node in $C^{(2)}_{r(i)}$, i.e., a hidden node anywhere in its activity period $T$.
\begin{eqnarray*}
\scriptstyle{R_i^{(3)}} &\scriptstyle{=}& \scriptstyle{\eta_i\Bigg(\prod_{j\in C^{(2)}_{r(i)}}\overline{h}_j\Bigg)\Bigg(1-\exp \Bigg\{-12~T_s\Bigg( \displaystyle{\sum_{j\in C^{(1)}_{r(i)}}}\overline{\tau}_j \Bigg) \Bigg\}.} \\
&& \scriptstyle{\exp \Bigg\{-T\Bigg(\displaystyle{\sum_{j\in C^{(2)}_{r(i)}}}\overline{\tau}_j\Bigg)\Bigg\}\Bigg) }
\end{eqnarray*}
\item[4.] The final term says that Node $i$ started its transmission as a simultaneous channel sensing with a node in $C^{(1)}_{r(i)}$ in the absence of any hidden node. This event surely ends up in a collision at the receiver $r(i)$.
\begin{eqnarray*}
R_i^{(4)} &=& \Bigg(\frac{\displaystyle{\sum_{j\in C^{(1)}_{r(i)}}\overline{\tau}_j}}{\beta_i+\displaystyle{\sum_{j\in \Omega_i}\overline{\tau}_j}}\Bigg)c_i\Bigg(\prod_{j\in C^{(2)}_{r(i)}}\overline{h}_j\Bigg) 
\end{eqnarray*}
\end{description}
All the other possible scenarios have a negligible probability. The overall packet collision probability is given by 
\begin{eqnarray}
p_i &=& \frac{R_i^{(1)}+R_i^{(2)}+R_i^{(3)}+R_i^{(4)}}{\eta_i+(1-\eta_i)c_i}									
\label{eqn:p_hidden}
\end{eqnarray}
The packet failure probability, $\gamma_i$, can now be calculated using Equations~\ref{eqn:p_hidden} and \ref{eqn:gamma}.

\subsection{Evaluation of Other Quantities}

\subsubsection{Average Service Rate ($\sigma_i$)}

Each packet that reaches the HOL position in the queue at a node can have multiple backoffs and transmissions before it is successfully transmitted or discarded. We define $\overline{Z}_i$ as the average time spent in backoff, and $\overline{Y}_i$ as the average transmission time until the packet is successfully transmitted or discarded at the MAC layer. Then the average service rate, $\sigma_i$ at node $i$ can be calculated as:
\begin{eqnarray}
\frac{1}{\sigma_i}=\overline{Z}_i + \overline{Y}_i
\label{eqn:sigma}
\end{eqnarray}
Using the default values from the standard, the mean backoff durations can be calculated (shown in Table~\ref{tab:mean_backoff_durations}).
\begin{table}[b]
\begin{center}
\vspace{-5mm}
{\scriptsize
\begin{tabular}{ccc}
\hline
\hline
\textbf{Backoff} & \textbf{Mean Backoff Duration} & \textbf{Mean Backoff Duration}\\
\textbf{Exponent} & \textbf{with successful CCA($T_s$)} & \textbf{with failed CCA($T_s$)}\\
\hline
\hline
$3$ & $70+20$ & $70+8$\\
\hline
$4$ & $150+20$ & $150+8$\\
\hline
$5$ & $310+20$ & $310+8$\\
\hline
\hline
\end{tabular}
}
\caption{Mean backoff durations in symbol times, $T_s$}
\label{tab:mean_backoff_durations}
\vspace{-5mm}
\end{center}
\end{table}
Define the following quantities:
\begin{eqnarray*}
\scriptstyle{\overline{B}_i}&\scriptstyle{=}&\scriptstyle{(70+20+158\alpha_i + 318\alpha_i^2 + 318\alpha_i^3 + 318\alpha_i^4 - 12\alpha_i^5)} \\
\scriptstyle{T_i^{(1)}}&\scriptstyle{=}&\scriptstyle{\Bigg(\frac{(70+20)(1-\alpha_i)}{(1-\alpha_i^5)}+\frac{248\alpha_i(1-\alpha_i)}{(1-\alpha_i^5)}+\frac{566\alpha_i^2(1-\alpha_i)}{(1-\alpha_i^5)}}\\
&&\scriptstyle{+\frac{884\alpha_i^3(1-\alpha_i)}{(1-\alpha_i^5)}+\frac{1202\alpha_i^4(1-\alpha_i)}{(1-\alpha_i^5)}\Bigg) }\\
\scriptstyle{T_i^{(2)}}&\scriptstyle{=}&\scriptstyle{(78+158+318+318+318)}
\end{eqnarray*}
where $\overline{B}_i$ refers to the mean backoff duration until the packet it transmitted or discarded due to successive CCA failures, $T_i^{(1)}$ has the interpretation of the mean backoff duration given that the packet transmission was successful, and $T_i^{(2)}$ is the mean time spent in backoff given that the packet was discarded due to successive CCA failures. Then the quantities $\overline{Z}_i$ and $\overline{Y}_i$ can be calculated as:
\begin{eqnarray*}
\scriptstyle{\overline{Z}_i}&\scriptstyle{=}&\scriptstyle{\alpha_i^5T_i^{(2)}+(1-\alpha_i^5)[T_i^{(1)}+\gamma_i(\alpha_i^5T_i^{(2)}+(1-\alpha^5)} \\
&&\scriptstyle{[T_i^{(1)}+\gamma_i(\alpha_i^5T_i^{(2)}+(1-\alpha_i^5)[T_i^{(1)}+\gamma_i(\alpha_i^5T_i^{(2)}+(1-\alpha_i^5)T_i^{(1)})])])]}\\
\scriptstyle{\overline{Y}_i}&\scriptstyle{=}&\scriptstyle{(1-\alpha_i^5)[T+\gamma_i(1-\alpha_i^5)[T+\gamma_i(1-\alpha_i^5)[T+\gamma_i(1-\alpha_i^5)T]]]}
\end{eqnarray*}

\subsubsection{Aggregate Arrival Rate ($\nu_i$), Goodput ($\theta_i$) and Discard Probability ($\delta_i$) for a Node}
\begin{figure}[h]
\begin{center}
	\includegraphics[scale=0.5]{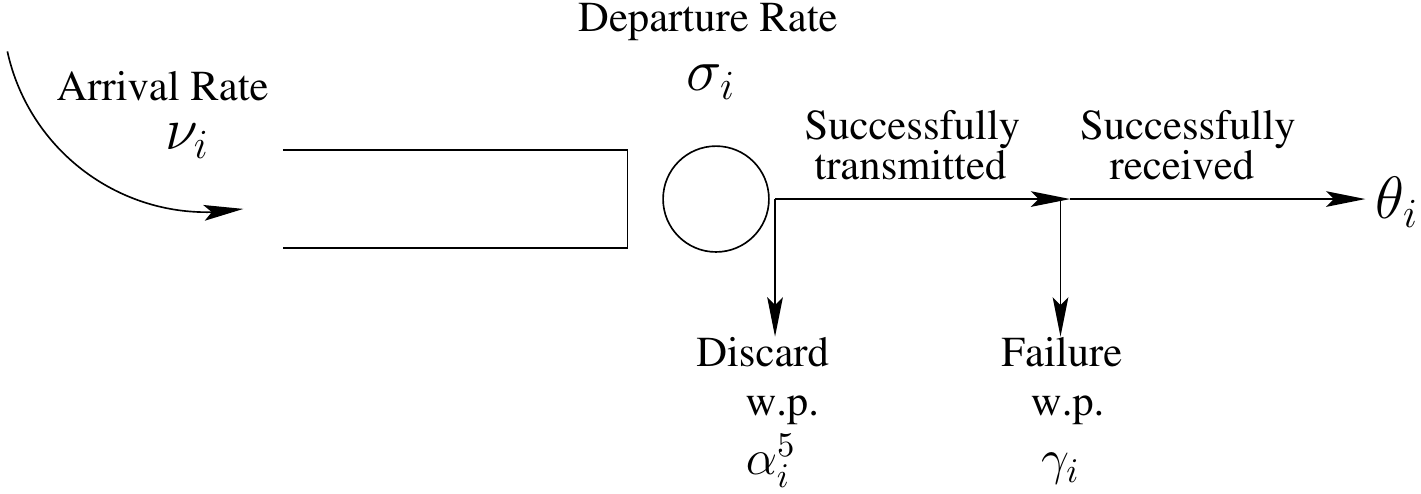}
	\caption{The goodput of a node is defined as the rate of successfully received packets by its receiver.}
	\label{fig:goodput}
	\vspace{-5mm}
\end{center}
\end{figure}

The arrival process at each node consists of packets which are generated at the same node (if it is a sensor) and the packets to be forwarded. The aggregate arrival rate at a node $i$ can be written as:
\begin{eqnarray}
\nu_i=\lambda_i+\sum_{k\in\mathcal{P}_i}\theta_k
\label{eqn:nu}
\end{eqnarray}
where $\lambda_i$ is the packet generation rate at sensor node $i$, $\mathcal{P}_i$ is the set of all its in-neighbours, and $\theta_i$ is the rate of packet transmission by the node which are \emph{successfully received} at $r(i)$ (known as goodput). An enqueued packet at a node is successfully received by the intended receiver in a manner shown in Figure~\ref{fig:goodput}.

Assuming that the queueing system is stable and has a steady-state solution, $\theta_i$ is given by
\begin{eqnarray}
\theta_i=\nu_i(1-\delta_i)
\label{eqn:theta_arrival}
\end{eqnarray}
where $\delta_i$ is the probability of discarding a packet due to consecutive CCA failures or successive failed retransmission attempts, and is given by
\begin{eqnarray}
\scriptstyle{\delta_i=\alpha_i^5+(1-\alpha_i^5)\gamma_i\Big[\alpha_i^5+(1-\alpha_i^5)\gamma_i\Big[\alpha_i^5+(1-\alpha_i^5)\gamma_i\Big[\alpha_i^5+(1-\alpha_i^5)\gamma_i\Big]\Big]\Big]}
\label{eqn:delta}
\end{eqnarray}
Note that if the queue at node $i$ is saturated, then the goodput $\theta_i$ is equal to $\sigma_i$. Note that while calculating $\sigma_i$, we have considered the packet discards into account. 

\subsubsection{The node non-empty probability, $q_i$}\label{non_empty_prob}
To find the expression for $q_i$, assuming that all the arriving packets reach the HOL position (i.e., no tail drops) and applying Little's Theorem, we get
\begin{eqnarray}
q_i=\frac{\nu_i}{\sigma_i}
\label{eqn:q_arrival}
\end{eqnarray}
Further, for a saturated node, the quantity $q_i$ is equal to $1$.

\subsubsection{Obtaining $b_i$ and $\beta_i$}

In order to find $b_i$, the fraction of time a node is in backoff provided it is non-empty, we embed a renewal process in conditional time where the renewal epochs are those instants at which the node enters the random backoff period after a packet transmission or packet discard. We use the RRT to find the expression for $b_i$ as 
\begin{eqnarray}
b_i=\frac{\overline{B}_i}{\overline{B}_i + (1-\alpha_i^5)T}
\label{eqn:calc_of_b}
\end{eqnarray}

\begin{figure}[h]
\begin{center}
\includegraphics[height=30mm, width=93mm]{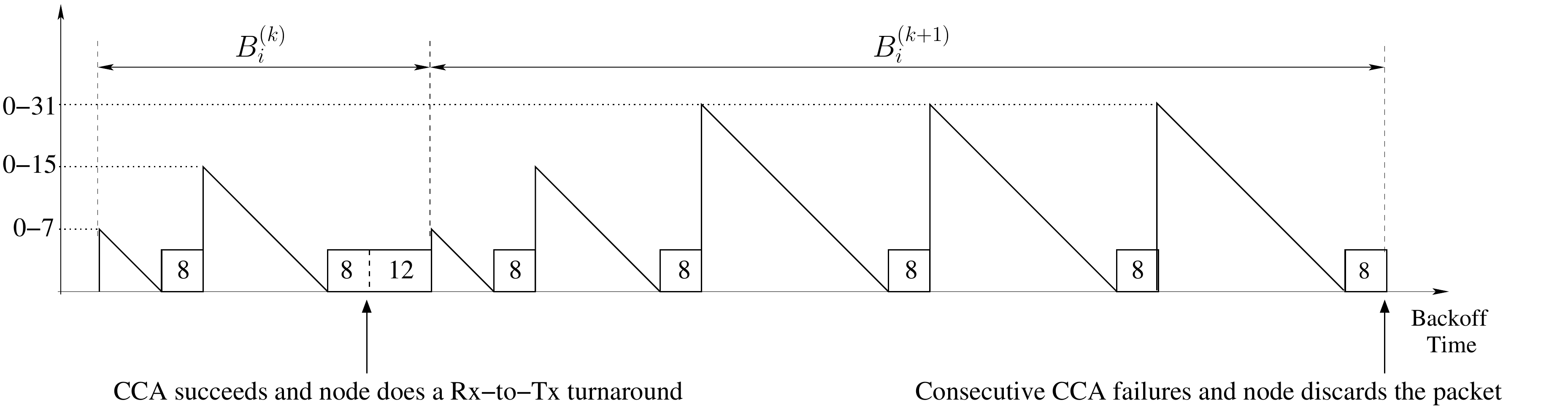}
\caption{Evolution of backoff periods in conditional (backoff) time. The $y$-axis is in units of a backoff slot ($20 ~T_s$). The range indicates that the random backoff duration is uniformly distributed within it.}
\label{fig:backoff_times}
\vspace{-5mm}
\end{center}
\end{figure}

The quantity $\beta_i$, the rate of CCA attempts in backoff times, is based on the backoff completion times irrespective of the transmission attempt of the packet since a packet retransmission is considered the same as a new packet transmission. Thus, for the calculation of $\beta_i$, it suffices to observe the process at node $i$ only in the backoff times. Figure~\ref{fig:backoff_times} shows the residual backoff process where after completion of the backoff duration, the node performs a CCA. 
Using the result of Kumar et al. \cite{winet.kumar-etal04new-insights} with their collision probability replaced by our CCA failure probability, $\alpha_i$, the expression for $\beta_i$ can be written as 
\begin{eqnarray}
\beta_i=\frac{1+\alpha_i+\alpha_i^2+\alpha_i^3+\alpha_i^4}{\overline{B}_i}
\label{eqn:beta}
\end{eqnarray}

\subsection{End-to-End Delay Calculation}

We use the two parameter Whitt's QNA \cite{winet.whitt83queueing-network-analyzer} to calculate the mean sojourn time at each node based on an approximate Moment Generating Function (MGF) of service time. We can write an expression for service time in a recursive manner by allowing infinite CCAs and retransmissions. To account for the discarded packets, we allow only a fraction of transmitted packets to join the next-hop neighbour's queue based on the discard probability at that node. We denote the service time at node $i$ by $S_i$ and let $B_i$ denotes the length of random backoff duration, which is assumed to have an exponential distribution with rate $\beta_i$ for node $i$. Then, $S_i$ can be written as
\begin{eqnarray*}
S_i &=& \begin{cases}
			B_i + \tilde{S}_i & \text{w.p. } \alpha_i \\
			B_i + T & \text{w.p. } (1-\alpha_i)(1-\gamma_i) \\
			B_i + T + \tilde{S}_i & \text{w.p. } (1-\alpha_i)\gamma_i
		\end{cases}
\end{eqnarray*}
where $\tilde{S}_i$ is a random variable with the same distribution as $S_i$. The MGF of $B_i$, denoted by $M_{B_i}(z)$, is equal to $\frac{\beta_i}{z+\beta_i}$. Therefore, we can express the MGF $M_{S_i}(z)$ of service time $S_i$ as 
\begin{eqnarray*}
M_{S_i}(z) &=& \frac{\beta_i(1-\alpha_i)(1-\gamma_i)e^{-zT}}{z+\beta_i(1-\alpha_i)(1-\gamma_ie^{-zT})}.
\end{eqnarray*}
The first two moments of the service time, $\mathbb{E}(S_i)$ and $\mathbb{E}(S_i^2)$ can be calculated by differentiating the MGF $M_{S_i}(z)$. The QNA procedure commences from the leaf nodes and converge to the base-station in a sequential manner. We first calculate the squared coefficient of variance of service time, denoted by $c_{S_i}^2$, at each node, i.e.,
\begin{eqnarray*}
c_{S_i}^2 &=& \frac{\mathbb{E}(S_i^2)}{(\mathbb{E}(S_i))^2} - 1
\end{eqnarray*}
Ignoring the packet discards at the MAC layer, the net arrival rate $\Lambda_i$ at a node $i$ is given by:
\begin{eqnarray*}
\Lambda_i &=& \begin{cases}
					\lambda_i + \displaystyle{\sum_{k\in \mathcal{P}_i}}\Lambda_k & \text{for source-cum-relay nodes} \\
					\displaystyle{\sum_{k\in \mathcal{P}_i}}\Lambda_k & \text{for relay nodes}
			  \end{cases}
\end{eqnarray*}
For any node $i$, let $\rho_i$ be defined as
\begin{eqnarray*}
\rho_i &=& \begin{cases}
					\lambda_i\mathbb{E}(S_i) & \text{for leaf nodes} \\
					\Lambda_i\mathbb{E}(S_i) & \text{for internal nodes}
			  \end{cases}
\end{eqnarray*}
We now calculate the squared coefficient of variance for the departure process ($c_{D_i}^2$) for a node which uses the squared coefficient of variance for the interarrival times at that node ($c_{A_i}^2$). Here we include the probability of a packet discard which is equal to $\delta_i$.
\begin{eqnarray*}
c_{D_i}^2 &=& 1 + \delta_i(\rho_i^2(c_{S_i}^2 - 1) + (1-\rho_i^2)(c_{A_i}^2 - 1))
\end{eqnarray*}
Further, $c_{A_i}^2$ is calculated as 
\begin{eqnarray*}
c_{A_i}^2 &=& \frac{1}{\Lambda_i}\Bigg(\lambda_i + \displaystyle{\sum_{j\in \mathcal{P}_i}}\Lambda_jc_{D_i}^2\Bigg)
\end{eqnarray*}
Finally, the mean sojourn time at a node $i$ is given by
\begin{eqnarray}
\Delta_i &=& \frac{\rho_i\mathbb{E}(S_i)(c_{A_i}^2 + c_{S_i}^2)}{2(1-\rho_i)} + \mathbb{E}(S_i)
\end{eqnarray}
The end-to-end mean packet delay for a source node $j$, provided that the set of nodes along the path from this node to the BS is $L_j$, is given by
\begin{eqnarray}
\overline{\Delta}_j &=& \displaystyle{\sum_{i\in L_j}\Delta_i}
\end{eqnarray}

\subsection{Packet Delivery Probability for each Source Node}
The fraction of packets generated at each source node that reaches the BS is called the packet delivery probability, denoted by $p_i^{(del)}$ for each source node $i$. Let the set of nodes constituting the path from a source node $i$ to the BS be $L_i$. Then, assuming that the drop events are independent from node to node, the expression for $p_i^{(del)}$ turns out to be
\begin{eqnarray}
p_i^{(del)} &=& \prod_{j\in L_i}(1-\delta_j)
\end{eqnarray}
where $\delta_j$ is the packet discard probability at node $j$.

\begin{figure}[t]
\begin{center}
	\includegraphics[scale=0.38]{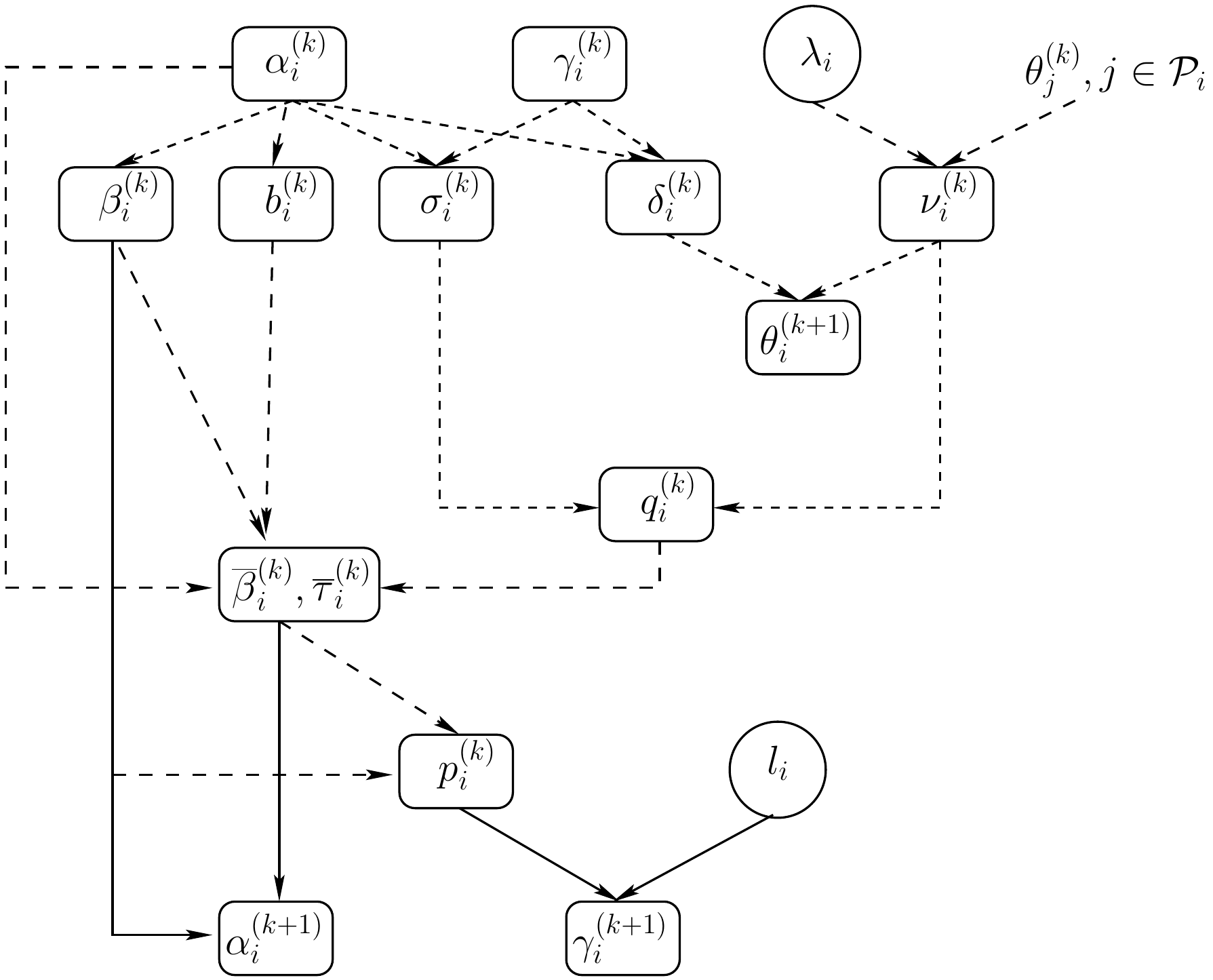}
	\caption{The global iteration scheme. The square boxes indicate the derived quantities and the round boxes indicate the known quantities.}
	\label{fig:iterations}
	\vspace{-8mm}
\end{center}
\end{figure}

\subsection{Global Iteration Scheme}\label{sec:iterations}
The global iteration scheme is shown in Figure~\ref{fig:iterations}. We start with a vector $\{0,0,\lambda_i\}$ corresponding to $\{\alpha_i,\gamma_i,\nu_i\}$ for each node $i$, and repeat the procedure until the quantities converge. 

\section{Numerical and Simulation Results} \label{sec:numerical-and-simulation-results}

\begin{figure}[p]
\begin{center}
\includegraphics[scale=0.35]{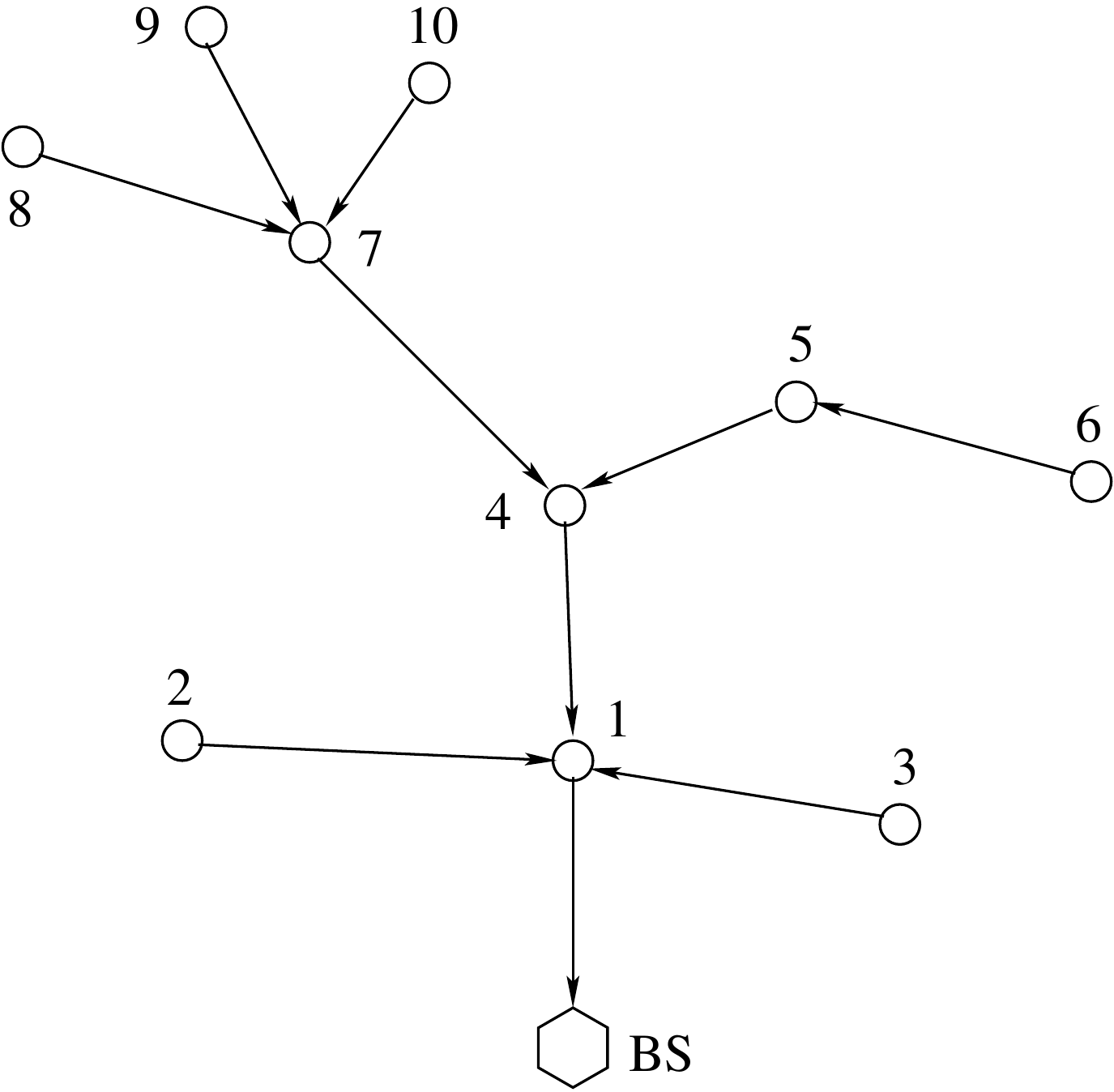}
\caption{A 10 nodes Tree topology. Note that each node can hear every other node.}
\label{fig:10nodes_tree}
\vspace{-10mm}
\end{center}
\end{figure}

\begin{figure}[p]
\begin{center}
	\includegraphics[scale=0.35]{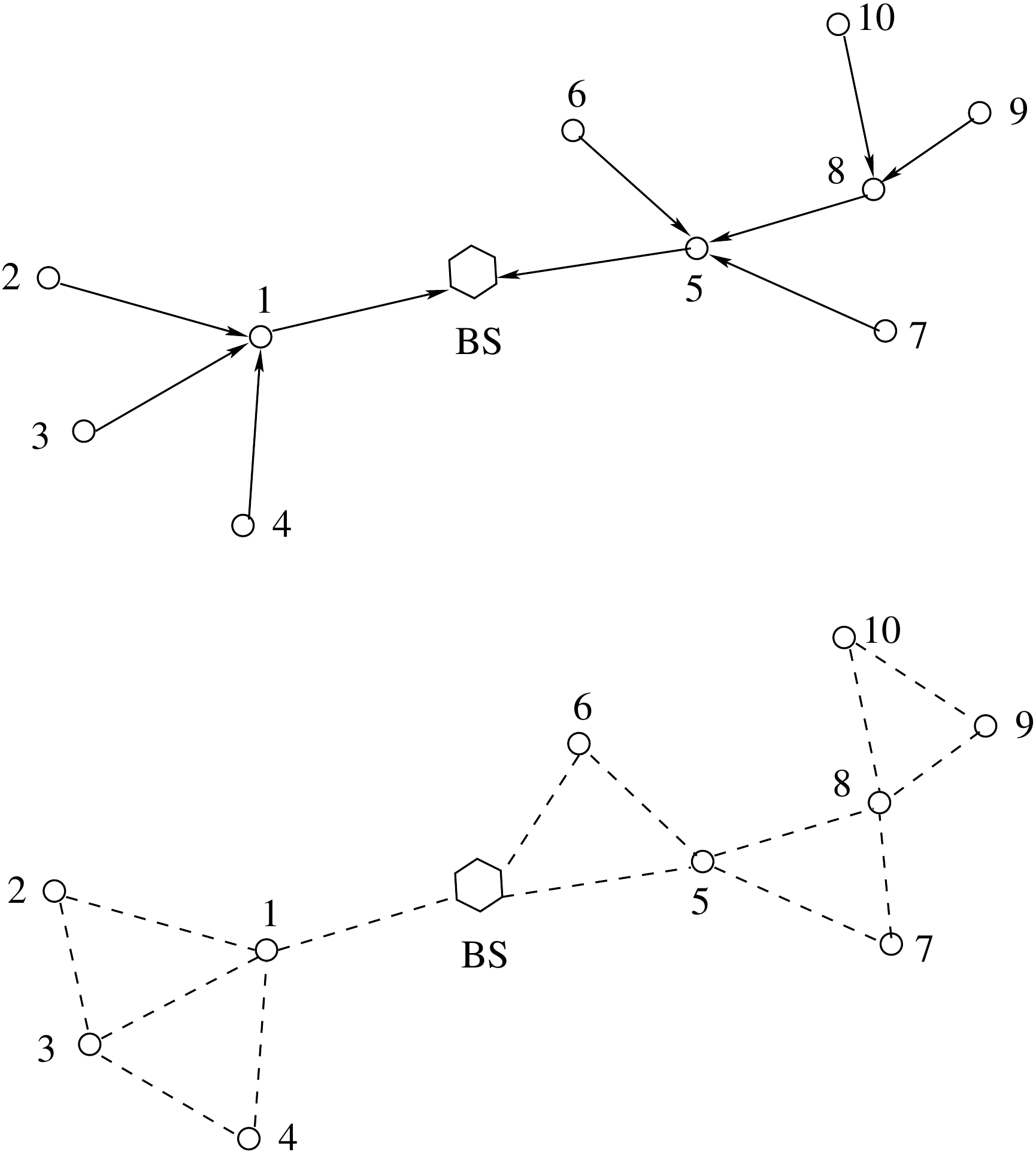}
	\caption{A 10 nodes TREE topology. Also shown is the dependency graph where the dotted lines connecting two nodes indicate that the nodes are in CS range of each other.}
	\label{fig:hidden_10nodes}
	\vspace{-8mm}
\end{center}
\end{figure}

For the verification of our analytical model, we use QualNet (v4.5) simulator with the default parameter values and a fixed packet size of $70$ bytes. However, the QualNet implementation is devoid of ACKs. Therefore, we compare results only for the ACK\emph{-less} scenarios (although the analysis is for a general scenario). The network shown in Figure~\ref{fig:10nodes_tree} has no hidden nodes whereas the one shown in Figure~\ref{fig:hidden_10nodes} has hidden nodes (the dependency graph is also shown). Note that all the nodes are sources (some of which also serve as relays) with identical packet generation rates that are simultaneously increased. 

In networks with no hidden nodes, we have accurate results for various quantities, viz., CCA failure probability ($\alpha_i$), packet discard probability ($\delta$), packet failure probability ($\gamma_i$), goodput ($\theta$) and node non-empty probability ($q$). Note that the discard probability is plotted on a log-log scale. It can be observed that the queue at node $1$ gets saturated when the packet generation rate reaches $20$ pkts/sec. However, the goodput $\theta_1$ starts to drop drastically after $5$ pkts/sec. Moreover, since node $5$ has a lower aggregare arrival rate than that of node $1$ ($\nu_5 < \nu_1$), its queue remains unsaturated even when $q_1$ becomes unity. The goodput $\theta_5$ also shows departure from the input rate $\nu_5$ at around $5$ pkts/sec. 

In the network with hidden nodes, the results for $\delta$, $\gamma$, $\theta$ and for $q$ match well. It can be seen that the goodput starts to deviate from the input rate at around $2.5$ pkts/sec for node $1$ and around $4$ pkts/sec for node $8$. There is a mismatch in CCA failure probability ($\alpha$) which is reflected in the plots for $b$, $\beta$, and $\Delta$. However, even for this scenario, the analysis captures the overall qualitative behavior. If the network is operated only until the drop probability is less than $10\%$, then the errors in all the measures for each node are less than $17\%$. 

Further, as observed for both the network types, the node packet discard probability increases to an impractical value before the average node delay becomes substantial, e.g., node $1$ discards $2$-$5$ packets for every $1000$ packets when the average node delay is $5$-$6$ msec. Hence, the overall delivery probabilities for the sources act as performance bottlenecks for these networks.   

\begin{figure}[p]
\begin{center}
	\includegraphics[height=7cm, width=9.5cm]{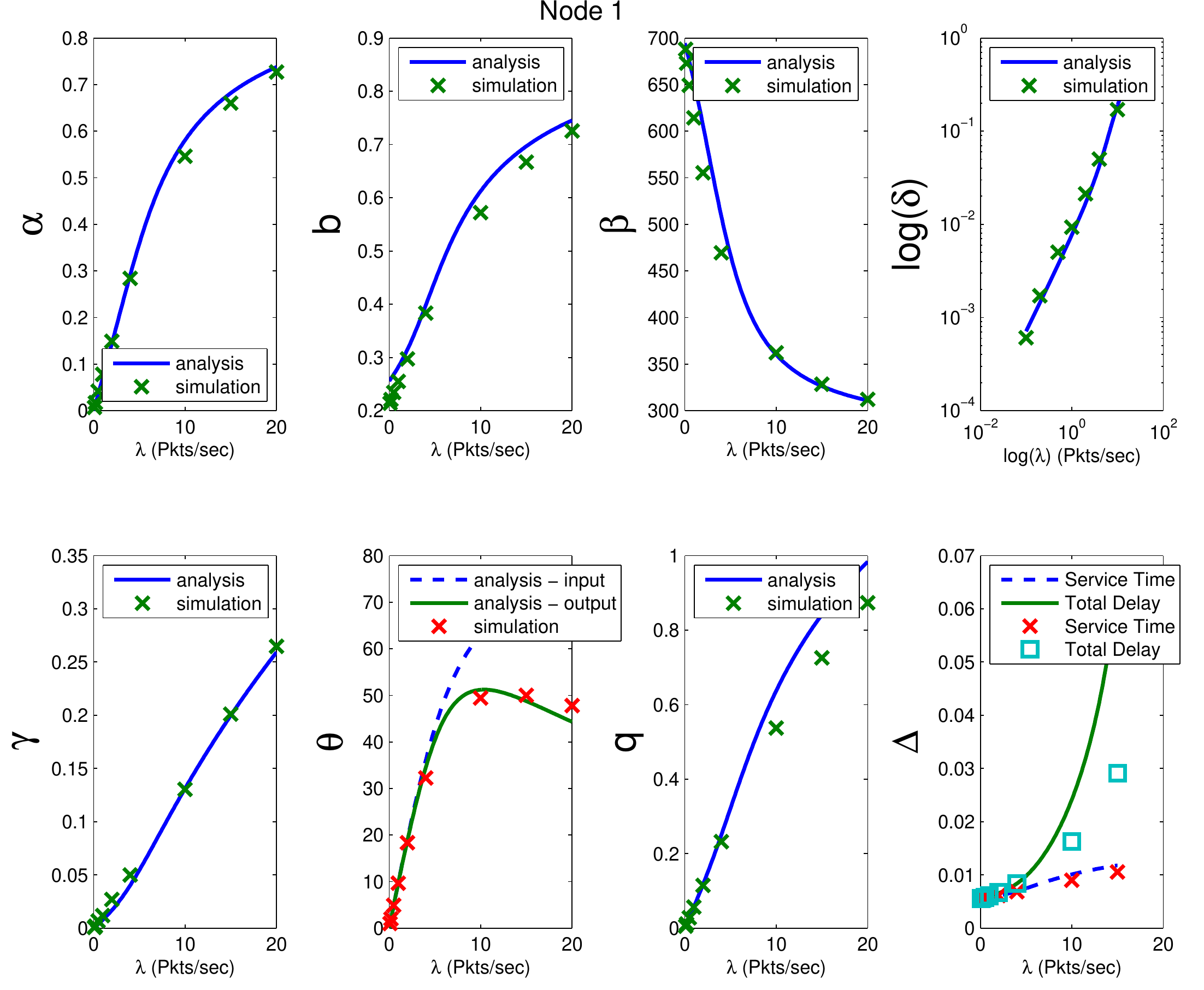}
	\caption{Results for Node 1 for the tree topology of Figure~\ref{fig:10nodes_tree}.}
	\label{fig:node1_tree}
\end{center}
\end{figure}

\begin{figure}[p]
\begin{center}
	\includegraphics[height=7cm, width=9.5cm]{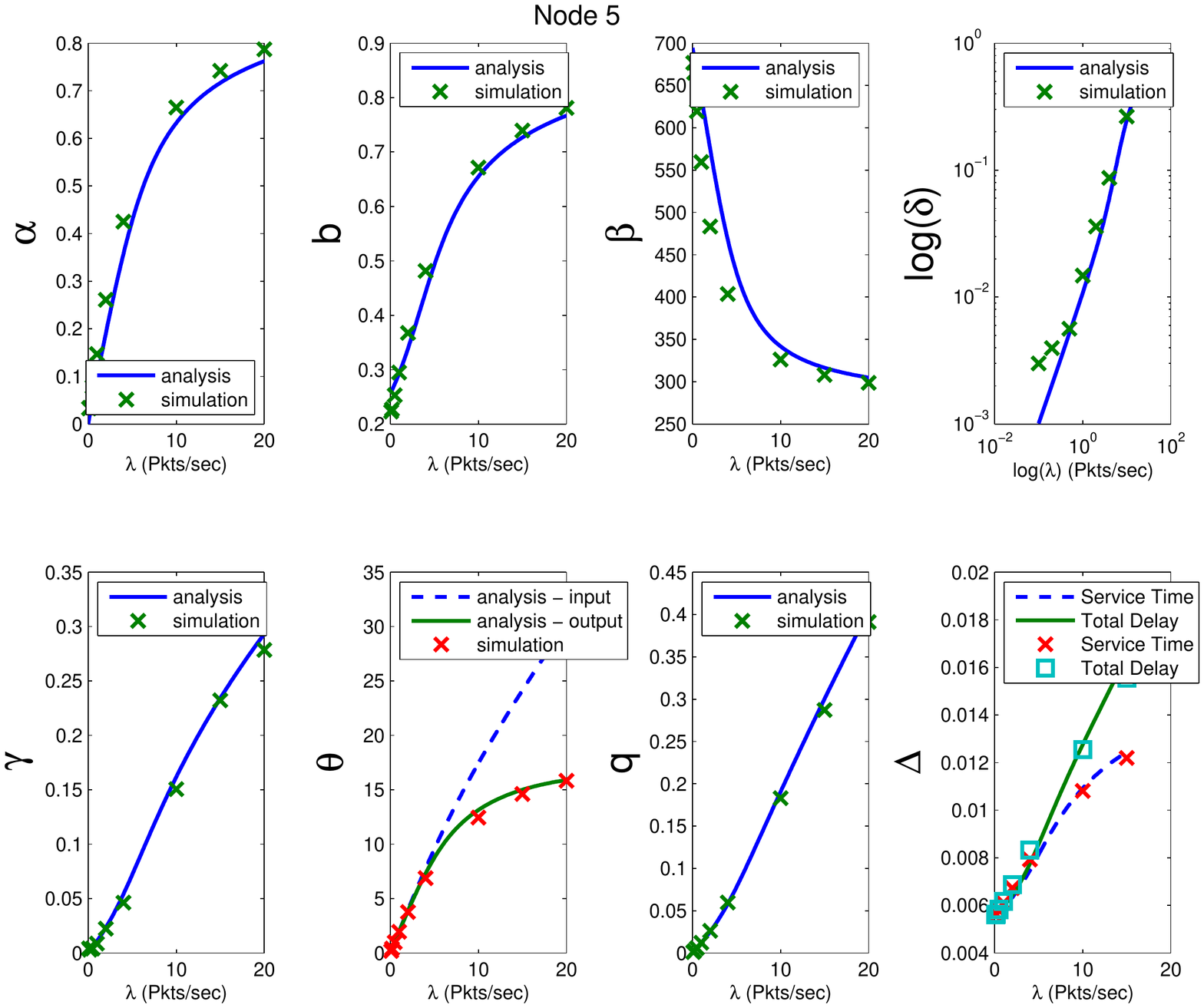}
	\caption{Results for Node 5 for the tree topology of Figure~\ref{fig:10nodes_tree}.}
	\label{fig:node5_tree}
\end{center}
\end{figure}

\begin{figure}[p]
\begin{center}
	\includegraphics[height=6cm, width=8cm]{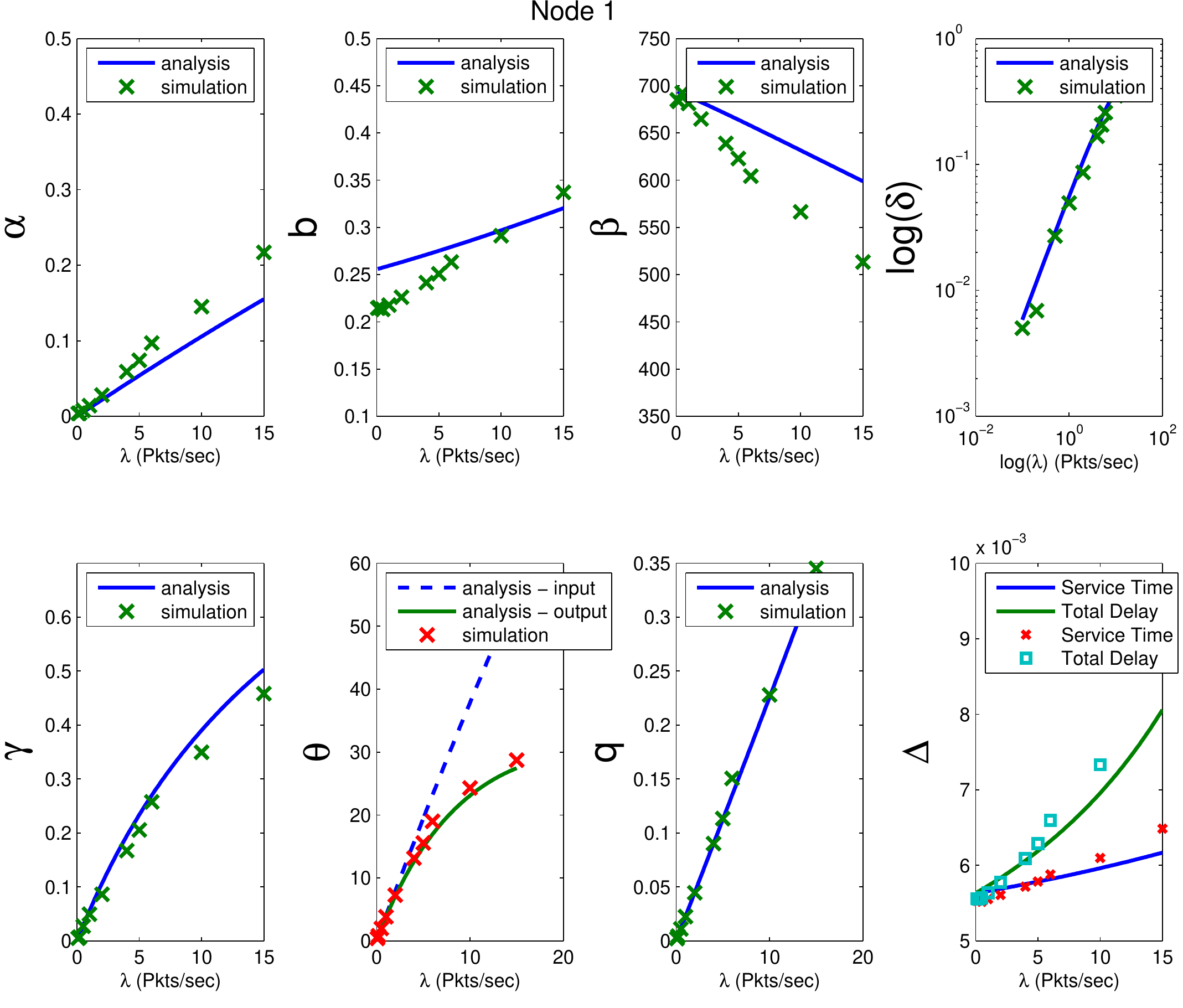}
	\caption{Results for Node 1 for the tree topology of Figure~\ref{fig:hidden_10nodes}.}
	\label{fig:hidden_10nodes_1}
\end{center}
\end{figure}

\begin{figure}[p]
\begin{center}
	\includegraphics[height=6cm, width=8cm]{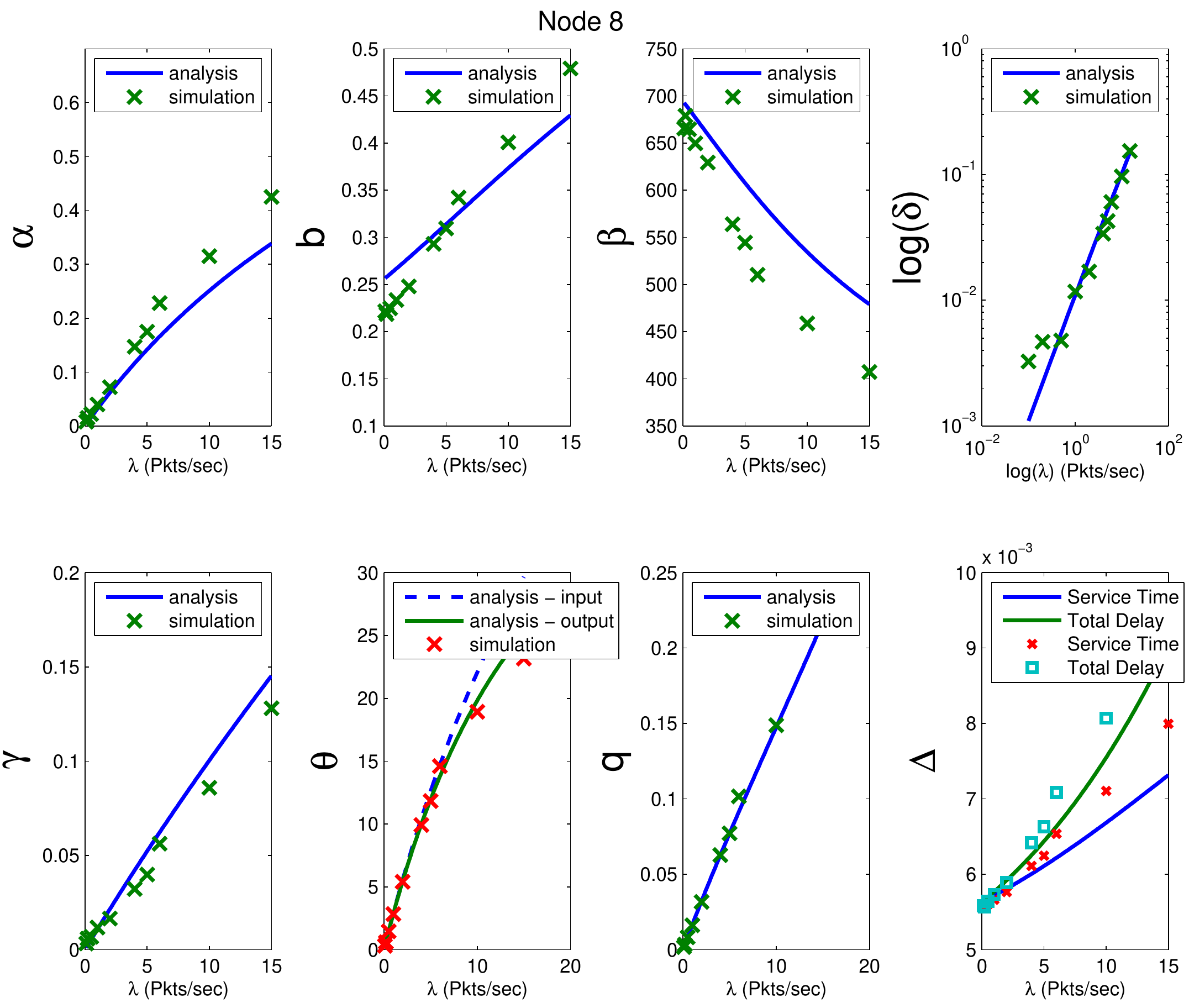}
	\caption{Results for Node 8 for the tree topology of Figure~\ref{fig:hidden_10nodes}.}
	\label{fig:hidden_10nodes_8}
\end{center}
\end{figure}

\section{Conclusion} \label{sec:conclusion}

We have developed an approximate stochastic model for the performance analysis of beacon-less \mbox{IEEE 802.15.4} multihop wireless networks, with arrivals. The model permits the estimation of several average performance measures. Our model is accurate for topologies devoid of hidden nodes and works reasonably well for topologies with hidden nodes in terms of the packet discard probability, failure probability and throughput. We calculated the mean end-to-end delays and packet delivery probabilities for each source in the network, which gives an idea of the QoS provided by the sensor network at a specific packet generation rate at source nodes. The results suggest that, for the relatively small size tree networks that we have studied, to operate in the low-discard low-delay region, the packet arrival rates at nodes should not be greater than a packet every few seconds (e.g. a packet inter-generation time of 5 to 10 seconds at the sources).

\bibliography{srivastava-kumar12beaconless-zigbee-analysis}

\end{document}